\documentclass[10pt,twocolumn,letterpaper]{article}

\usepackage{amsfonts}
\usepackage{amsmath}
\DeclareMathOperator{\Tr}{tr}
\usepackage{mathabx}					 	
\usepackage{placeins}                       
\usepackage{subcaption,caption}
\usepackage{epsfig}
\usepackage{booktabs}                       
\usepackage{tabularx}                       
\graphicspath{{Figures/}}                   
\usepackage{accents}                        
\usepackage{supertabular}
\usepackage{multirow}
\usepackage{longtable}
\usepackage{capt-of}                        
\usepackage{algorithm}                   
\usepackage{algpseudocode}
\usepackage[hyphens]{url}
\usepackage[colorlinks=true,urlcolor=blue]{hyperref} 

\usepackage{iccv}
\usepackage{times}
\usepackage{graphicx}
\usepackage{amssymb}
\usepackage{bbm}
\usepackage{mathrsfs}

\iccvfinalcopy 

\def\iccvPaperID{406} 

\begin{document}

\title{Multi-objective analysis of the Sand Hypoplasticity model calibration}

\author{Francisco J. Mendez{$^{1,*}$},~ Miguel A. Mendez{$^{2}$}, ~ Nicola Sciarra{$^{1}$}~, Antonio Pasculli{$^{1}$}\\
{$^{1}$}{\small University G. D'Annunzio, Dept. of Engineering Geology (INGEO), Chieti-Pescara, Italy};\\ {$^{2}$} {\small von Karman Institute for Fluid Dynamics, EA Department, Sint-Genesius-Rode, Belgium}\\
{\small{{$^{*}$}Corresponding to: \tt{francisco.mendez@unich.it}}}}


\maketitle
\thispagestyle{empty}

\begin{abstract}
The Sand Hypoplastic (SH) constitutive law by von Wolffersdorff (1996) is a interesting hypoplastic model for soil mechanics. This model includes eight parameters, usually calibrated using the oedometric (OE) and the drained isotropically consolidated triaxial tests (CD). However, previous studies show that the SH model calibration in the CD test has conflicting requirements in predicting the evolution of stresses and strains.

In this work, we study the SH model calibration over a wide range of testing conditions using 12 OE and 25 CD tests by Wichtmann and Triantafyllidis (2016) on the Karlsruhe sands. The parameter space is extensively explored via Genetic Algorithm Optimization (GA) using the recently developed open-source software GA-cal (available at \url{https://github.com/FraJoMen/GA-cal}). This exploration allowed us to study the SH model's predictive limits and to identify, using a multi-objective analysis, the main parameters governing the compromise between the accurate prediction of stresses versus strain in the CD tests.

\end{abstract} 

\textbf{Keywords}~~Automatic calibration, Hypoplasticity, Numerical optimization, Genetic Algorithm, Nonlinear Regression

\section{Introduction}
Based on the first Hypoplastic constitutive law \cite{Wei_Kolymbas_Numerical_testing_of_the_stability_criterion_for_hypoplastic_constitutive_equations}, several models have been developed to describe the stress-strain behaviour of clay \cite{Wang2020,Wang_2020,Masin_Clay_hypoplasticity_model_including_stiffness_anisotropy,Masin_A_hypoplastic_constitutive_model_for_clays,Masin_Clay_hypoplasticity_with_explicitly_defined_asymptotic_states} and sand \cite{Wu_A_basic_hypoplastic_constitutive_model_for_sand,Fuentes2012,WU199645,Gudehus_A_comprehensive_,Bauer_Calibration_of_a_comprehensive_hypoplastic_model_for_granular_materials}, with extensions available for cyclic loads  \cite{Niemunis_Herle__Hypoplastic_model_for_cohesionless_soils_with_elastic_strain_range,Fuentes_and_Triantafyllidis,Tafili2020,Fuentes2020}. Arguably one of the most famous hypoplastic model is the one proposed by von Wolffersdorff \cite{Wolffersdorff_A_hypoplastic_for_granular_material_with_a_predefined_limit_state_surface}, often referred to as Sand Hypoplasticity (SH). This model has been successfully applied to a wide variety of problems in geotechnical engineering \cite{WangChengwu2022,MACHACEK2021104276,Staubach_contact},  despite its limitations \cite{Wu_A_basic_hypoplastic_constitutive_model_for_sand}.


The SH model requires eight parameters. These are traditionally calibrated from geotechnical laboratory tests, as firstly proposed in \cite{Herle_Gudehus_Determination_of_parameters_of_a_hypoplastic_constitutive_model_from_properties_of_grain_assemblies}. The early approaches for the SH model calibration, however, relied on a large number of laboratory tests and estimates of the tangents to the response curves. Unfortunately, such estimates are usually impossible in some tests (e.g. in the oedometric curves) because of the coarse time-stepping, and are particularly sensitive to measurement noise in others (e.g. in the triaxial tests). 
Moreover, the formulas used to compute the model parameters are extremely sensitive to the input data, resulting in considerable uncertainties in the estimated parameters and the need for non-trivial manual adjustments to improve the model predictions.

To reduce the manual adjustments in the SH calibration, a more robust approach has been proposed in \cite{Kadlicek2022_I,Kadlicek2022_II,Kadlicek_Calibration_of_Hypoplastic_Models_for_Soils,Kadlicek_Automatic_online_calibration_software_excalibre} and made available in the free software ExCalibre (\url{www.soilmodels.com/excalibre}). This approach solely relies on experimental data from the oedometric (OE) and drained isotropically consolidated triaxial tests (CD) tests and uses a combination of empirical relations and optimization. Interestingly, this approach ignores the SH model predictions of the volumetric strain and uses the OE data to calibrate some of the parameters and the CD data to calibrate others. This is a convenient but restrictive simplification because all parameters play a role in determining the soil response in both tests.

More general methods formulating the calibration of constitutive laws as a regression problem were proposed in \cite{ZENTAR2001,Yazdani2013,Yin_2018}. The regression solution involves an optimization, for which both deterministic and stochastic approaches have been proposed, as reviewed in \cite{Kadlicek2022_I} and \cite{Kadlicek2022_II}. 

Stochastic Optimisers (SO) are notoriously less sample efficient than deterministic ones (see \cite{Papon_2012}) but allow for better exploring the parameter space without being trapped in local minima. These optimisers are based on the generation of pseudorandom numbers commonly used by the  Monte Carlo method, which has also been applied in different fields in \cite{Calista_PASCULLI,PASCULLI2018370,PasculliPugliese,MoniaPasculli}. The most popular SO used in the parameter calibration of a constitutive model are the genetic algorithm (GA) \cite{PAL1996325,JAVADI19991,doi10106140771,Yin2017,Schorr}, the differential evolution  \cite{VARDAKOS2012109,DEA_optimization,Machacek2022}, and particle swarm optimisation \cite{LI2022,BHARAT2009984,ZHANG2009604}. The authors recently proposed the first calibration of SH using GA in \cite{Mendez2021}. In this work, the GA was proven able to provide a good calibration and was also employed to quantify the model prediction's sensitivity, the model parameters' uncertainty and the potential correlations between them.

The availability of powerful optimization techniques and ever-increasing experimental datasets allows for exploring and stretching the SH model's predictive capabilities. To the author's knowledge, a practical problem that has yet to be addressed in the open literature is the extent to which the SH model can perform well in both OE and CD tests (thus, the calibration is possible) for different types of sands. While the SH model in the first calibration on the Hochstetten's sand by von Wolffersdorff \cite{Wolffersdorff_A_hypoplastic_for_granular_material_with_a_predefined_limit_state_surface,Mendez2021} performed satisfactorily in all tests, many authors have reported on the limits of this model in predicting the volumetric deformations in the CD tests for other sands \cite{Wichtmann2016-xw,Kadlicek2022_II,Machacek2022}. In this work, we study these limits and the conflicting requirements on the model parameter to predict stresses and strain in the CD test.

Building upon an extensive exploration of the parameter space via GA, this work aims to investigate the SH model capabilities to describe the 12 OE and 25 CD tests\footnote{The data set is downloadable from the website \url{www.torsten-wichtmann.de/}} on the Karlsruhe sands published in \cite{Wichtmann2016-rs}. Moreover, with the adaptive weighted sum approach to multiobjective optimization \cite{Kim2006}, we analyze the relationship between the SH model parameters and the model performances in the CD tests. The analysis was carried out using the software GA-cal, an open-source software recently released by the authors \cite{Mendez_GA-cal_2022} for solving the inverse problem of calibrating a constitutive law. The source code of GA-cal, a Fortran revisited version of the algorithms initially developed in Python \cite{Mendez2021}, is freely available at \url{https://github.com/FraJoMen/GA-cal}.

The article is organized as follows. Section 2 recalls the calibration problem, while Section 3 describes the dataset and the preprocessing. Section 4 presents the calibration results and Section 5 collects the conclusions.

\section{The SH Calibration Problem}\label{sec2}

The SH model proposed by \cite{Wolffersdorff_A_hypoplastic_for_granular_material_with_a_predefined_limit_state_surface} links the time derivative of the Cauchy effective stress $\dot{\mathbf{T}}$ and the void ratio $\dot{e}$, with the granulate stretching rate $\mathbf{D}$ and the state variable $e$ and $\mathbf{T}$.
This link depends on eight parameters; hence the SH model in the OE and CD test case can be written as a system of parametric nonlinear Ordinary Differential Equations (ODEs) as 

\begin{equation}
    \label{ODES}
    \dot{\mathbf{X}}=\mathbf{H}(\mathbf{X}; \mathbf{P})\,;\,\mathbf{X}(0)=\mathbf{X}_0\,,
\end{equation} where $\mathbf{X}=[\mathbf{T};e]$ is the state vector and $\mathbf{P}$ is the set of parameters:

\begin{equation}
  \mathbf{P}=\{e_{c0} ,  e_{d0} ,  e_{i0} ,  h_s ,  \varphi_c ,  n ,  \alpha ,  \beta \}\,.
\end{equation} The reader is referred to appendix \ref{App} for more details on this model.  

For the scopes of this work, it suffices to recall that the model calibration consists in identifying the parameters $\mathbf{P}$ that make the trajectory of the system $\mathbf{X}(t)$ match with a set of experimental data $\mathbf{X}(t_k)$ available on a discrete set of times $k=1,\dots n_t$.

The experimental results are given in the plane $(\sigma_v$,\,$e$) for the OE test, with $\sigma_v$ the effective vertical pressure, and in the triaxial deviatoric $(\varepsilon_a,\,q$) and triaxial volumetric  $(\varepsilon_a,\,\varepsilon_v$) planes for the CD test, with $q$ the deviatoric stress, and $\varepsilon_a$ and $\varepsilon_v$ the volumetric and the axial deformations.

Following the approach proposed in \cite{Mendez2021}, the calibration aims to find the parameters that minimise a cost function. This cost function $C(\mathbf{P})$ measures the discrepancy between model prediction and available data. As in \cite{Mendez2021}, we define it as:

\begin{equation} \label{eq:cost} C(\mathbf{P})=w_1\delta_1(\mathbf{P})+w_2\delta_2(\mathbf{P})+w_3\delta_3(\mathbf{P})\,,
\end{equation} where $w_i$ weights the discrepancy in each plane on the model performance evaluation $(\sigma_v$,\,$e$), $(\varepsilon_a,\,q$) and $(\varepsilon_a,\,\varepsilon_v$). The $\delta_i$, with $i=1,2,3$ functions in the \eqref{eq:cost} are the average Fréchet distances between data and predictions, measured in appropriate dimensionless plans (see \cite{Mendez_GA-cal_2022}). 

The weight in the \eqref{eq:cost} can be used to prioritise the model performance on one plane or the other. In this work, we use these values to study the feasibility of identifying a set performing well on \emph{all} planes and to reveal contrasting requirements.

Using the GA-cal code, we minimise \eqref{eq:cost} and explore the parameter space (i.e. the set of possible parameters) with Genetic Algorithm (GA) optimisation.

\section{Dataset}\label{sec:Dataset}

The selected experimental data were collected on the \textit{Karlsruhe fine sand}. This sand is characterised by the subangular shape of most grains and no fines content, a mean grain size $d_{50} = 0.14$ mm and a uniformity coefficient $C_u = d_{60}/d_{10} = 1.5$  \cite{Wichtmann2016-rs}.  

The initial conditions for the OE tests are reported in Table \ref{tab:OED_init}, while the initial conditions for the CD tests are reported in Table \ref{tab:TxD_in}. These tables also report the initial relative densities of the samples $I_{D0}$.

\begin{equation}
 I_{D0} = \frac{e_{max} - e_0}{e_{max}- e_{min}}\,.   
\end{equation} The minimum and maximum void ratios $e_{min} = 0.677$ and $e_{max} = 1.054$ were determined from standard tests at mean pressure $p = 0$ \cite{Wichtmann2016-rs}. 
\begin{table}[tbp]
  \centering
  \caption{Oedometric compression tests. The void ratios
$e_0$ and relative densities $I_{D0}$ are measured at the initial conditions.}
    \begin{tabular}{lrr}
    \toprule
    Test  & \multicolumn{1}{l}{$e_0$} & \multicolumn{1}{l}{$I_{D0}$} \\
    No.   & \multicolumn{1}{l}{[-]} & \multicolumn{1}{l}{[-]} \\
    \midrule
    OE1   & 1.039 & 0.04 \\
    OE2   & 1.029 & 0.07 \\
    OE3   & 0.99  & 0.17 \\
    OE4   & 0.971 & 0.22 \\
    OE5   & 0.946 & 0.28 \\
    OE6   & 0.908 & 0.39 \\
    OE7   & 0.846 & 0.55 \\
    OE8   & 0.833 & 0.59 \\
    OE9   & 0.808 & 0.65 \\
    OE10  & 0.777 & 0.73 \\
    OE11  & 0.74  & 0.83 \\
    OE12  & 0.721 & 0.88 \\
    \bottomrule
    \end{tabular}%
  \label{tab:OED_init}%
\end{table}%

\begin{table}[htbp]
  \centering
  \caption{Drained monotonic triaxial compression
tests. The void ratios $e_0$ and relative densities $I_{D0}$ are measured at the initial mean pressure $p_0$, prior to shearing.}
    \begin{tabular}{rrrr}
    \toprule
    \multicolumn{1}{c}{Test} & \multicolumn{1}{c}{$e_0$} & \multicolumn{1}{c}{$I_{D0}$} & \multicolumn{1}{c}{$p_0$} \\
    \multicolumn{1}{c}{No.} & \multicolumn{1}{c}{[-]} & \multicolumn{1}{c}{[-]} & \multicolumn{1}{c}{[kPa]} \\
    \midrule
    CD1     & 0.996 & 0.15  & 50 \\
    CD2     & 0.975 & 0.21  & 100 \\
    CD3     & 0.975 & 0.21  & 200 \\
    CD4     & 0.97  & 0.22  & 300 \\
    CD5     & 0.96  & 0.25  & 400 \\
    CD6     & 0.88  & 0.46  & 50 \\
    CD7     & 0.862 & 0.51  & 100 \\
    CD8     & 0.859 & 0.52  & 200 \\
    CD9     & 0.848 & 0.55  & 300 \\
    CD10    & 0.847 & 0.55  & 400 \\
    CD11    & 0.84  & 0.57  & 50 \\
    CD12    & 0.819 & 0.63  & 100 \\
    CD13    & 0.824 & 0.63  & 200 \\
    CD14    & 0.822 & 0.64  & 300 \\
    CD15    & 0.814 & 0.68  & 400 \\
    CD16    & 0.743 & 0.82  & 50 \\
    CD17    & 0.758 & 0.79  & 100 \\
    CD18    & 0.748 & 0.81  & 200 \\
    CD19    & 0.734 & 0.85  & 300 \\
    CD20    & 0.753 & 0.8   & 400 \\
    CD21    & 0.734 & 0.85  & 50 \\
    CD22    & 0.735 & 0.85  & 100 \\
    CD23    & 0.706 & 0.92  & 200 \\
    CD24    & 0.697 & 0.95  & 300 \\
    CD25    & 0.718 & 0.89  & 400 \\
    \bottomrule
    \end{tabular}%
  \label{tab:TxD_in}%
\end{table}%

\subsection{Pre-processing and Test Grouping}

To speed up the cost function evaluation (and hence allow for better exploration of the parameter space), the calibration is not carried out on the full data set. Instead, a sub-sampling was carried out by interpolating the available data. 10 points were uniformly distributed between 10 and 400 kPa for the OE test, while 15 points were uniformly distributed between points between $\varepsilon_a = 0.01 \% $ and $max(\varepsilon_a) $. 

The significant sub-sampling drastically reduced the computational time with a negligible loss in accuracy, given the model predictions' smoothness.

To analyse the robustness of the model calibration, the large number of experimental tests were divided into six groups characterised by different initial relative densities $I_{D0}$. Among the initial conditions identifying each test, the initial void ratio was found to be the one that most influenced the model calibration. The different groups, labelled G1, G2,..., G6 are listed in Table \ref{tab:GA_grup}. These also differ in the number of tests.

\begin{table}[htbp]
  \centering
  \caption{Groups of experimental data were used in the calibrations.}
    \begin{tabular}{lcl}
    \toprule
     Groups labels  & Tests    & Tests numbers  \\
     \midrule
    \multirow{2}[1]{*}{G1} & OE    & 1, 2, ..., 12 \\
          & CD    & 1, 2, ..., 25  \\
    \midrule
    \multirow{2}[2]{*}{G2} & OE    & 3, 4 \\
          & CD    & 1, 2, ..., 5 \\
    \midrule
    \multirow{2}[2]{*}{G3} & OE    & 6, 7 \\
          & CD    & 6, 7, ..., 10 \\
    \midrule
    \multirow{2}[2]{*}{G4} & OE    & 8, 9  \\
          & CD    & 11, 12, ..., 15 \\
    \midrule
    \multirow{2}[2]{*}{G5} & OE    & 10, 11 \\
          & CD    & 16, 17, ..., 20 \\
    \midrule
    \multirow{2}[2]{*}{G6} & OE    & 11, 12 \\
          & CD    & 21, 22, ..., 25 \\
    \bottomrule
    \end{tabular}%
  \label{tab:GA_grup}%
\end{table}%

All input files, output files and the Python scripts for post-processing used in this work are made available in the folder \textit{Example} in the GA-cal repository \url{https://github.com/FraJoMen/GA-cal}.

\section{Results}\label{sec:Result}

This section is divided into two parts. In the first (Sec. \ref{R1}), we analyse the GA calibrations on the previously defined groups to study the range of validity of each calibration and its sensitivity to the available data. In the second part (Sec. \ref{R2}), we report on the Pareto front analysis to identify which parameters mostly influence the model performances in the different tests.

\subsection{The variability of SH model calibration }\label{R1}

For each of the six groups of calibration data sets in Table \ref{tab:GA_grup}, we consider four sets of weights $(w_1,w_2,w_3)$ (see equation \eqref{eq:cost}). These are $(1,1,1)$,$(1,0,0)$,$(0,1,0)$,$(0,0,1)$; that is, the first gives equal importance to all planes, the second only focuses on the $(\sigma_e,e)$ plane, the third on the $(\varepsilon_a,q)$ plane, etc.

Table \ref{tab:PARAMETI_} reports a set of 24 possible combinations of parameters obtained via GA (labelled as C02, C03, etc) together with the parameters indicated by Wichtman and Triantafyllidis (WT) \cite{Wichtmann2016-rs} (labelled as C01) for reference. For each combination, the table reports the weights used in the cost function and the data group used in the calibration. Moreover, the last column collects the fitness score in \eqref{eq:cost}, weighted by the number of tests $n_i$ in the group used for the calibration and the sum of weights:

\begin{equation}\label{eq:eta}
\eta= \frac{C(\mathbf{P})}{\sum^{3}_{j=1} w_j\, n_i}
\end{equation} This normalization enables a fair comparison between parameters that have been evaluated in a different number of planes.

The values of $\eta$ show that the parameters derived by the GA calibration on all the data (combination C02-C07) outperform the parameter sets by Wichtman and Triantafyllidis (C01) according to the metrics in \eqref{eq:cost}. The error significantly reduces when the calibration focuses on one of the planes. Moreover, the test in group G2 apparently yields the largest error in the $(\varepsilon_a,\,\varepsilon_v)$ plane.

\begin{table*}[t]
  \centering
  \caption{Summary of the optimal (calibrated) SH model parameter using GA-cal (C02,C03,$\dots$,C25), together with the ones identified by Wichtman and Triantafyllidis (C01). The table also collects the weights $w_i$, $i=1,2,3$ and the group (see Table \ref{tab:GA_grup}) used for the calibration as well as the fitness score $\eta$ (see equation \ref{eq:eta}).}
    \begin{tabular}{cccccrrrrrrrrr}
    \toprule
    \multirow{3}{*}{ID} & \multirow{3}{*}{$w_1$} & \multirow{3}{*}{$w_2$} & \multirow{3}{*}{$w_3$} & \multirow{3}{*}{Group} & \multicolumn{1}{c}{$\varphi_c$} & \multicolumn{1}{c}{$h_s$} & \multicolumn{1}{c}{$n$} & \multicolumn{1}{c}{$e_{d0}$} & \multicolumn{1}{c}{$e_{c0}$} & \multicolumn{1}{c}{$e_{i0}$} & \multicolumn{1}{c}{$\alpha$} & \multicolumn{1}{c}{$\beta$} & \multicolumn{1}{c}{$\eta$}\\
    \cmidrule(r){6-14}
         &  &  &  &  & \multicolumn{1}{c}{(°)} & \multicolumn{1}{c}{(GPa)} & \multicolumn{1}{c}{(-)} & \multicolumn{1}{c}{(-)} & \multicolumn{1}{c}{(-)} & \multicolumn{1}{c}{(-)} & \multicolumn{1}{c}{(-)} & \multicolumn{1}{c}{(-)} & \multicolumn{1}{c}{(\%)}\\
    \midrule
    C 01     & -     & -     & -     & - & 33.10 & 4.00  & 0.27  & 0.672 & 1.050 & 1.208 & 0.14  & 2.50 & 2.09 \\
    C 02     & 1     & 1     & 1     & G1 & 32.58 & 4.03  & 0.28  & 0.636 & 1.060 & 1.219 & 0.28  & 1.50 & 1.21 \\
    C 03     & 1     & 1     & 1     & G2 & 32.94 & 2.27  & 0.29  & 0.621 & 1.070 & 1.220 & 0.18  & 1.33 & 1.02 \\
    C 04     & 1     & 1     & 1     & G3 & 32.91 & 4.18  & 0.30  & 0.616 & 1.080 & 1.231 & 0.20  & 1.44 & 0.94 \\
    C 05     & 1     & 1     & 1     & G4 & 31.97 & 2.94  & 0.31  & 0.626 & 1.080 & 1.188 & 0.29  & 1.30 & 0.88\\
    C 06     & 1     & 1     & 1     & G5 & 31.69 & 3.83  & 0.30  & 0.626 & 1.080 & 1.210 & 0.27  & 1.38 & 0.78\\
    C 07     & 1     & 1     & 1     & G6 & 31.75 & 4.63  & 0.30  & 0.621 & 1.070 & 1.231 & 0.28  & 1.34 & 0.71\\
    C 08     & 1     & 0     & 0     & G1 & 35.36 & 6.90  & 0.27  & 0.557 & 0.960 & 1.142 & 0.28  & 1.82 & 0.66\\
    C 09     & 1     & 0     & 0     & G2 & 29.75 & 7.86  & 0.24  & 0.559 & 0.980 & 1.215 & 0.29  & 1.77 & 0.31\\
    C 10    & 1     & 0     & 0     & G3 & 35.37 & 5.33  & 0.30  & 0.545 & 0.940 & 1.137 & 0.28  & 1.83 & 0.32\\
    C 11    & 1     & 0     & 0     & G4 & 25.32 & 7.50  & 0.30  & 0.553 & 0.970 & 1.213 & 0.10  & 1.17 & 0.52\\
    C 12    & 1     & 0     & 0     & G5 & 29.04 & 7.54  & 0.25  & 0.614 & 1.040 & 1.300 & 0.27  & 1.71 & 0.24\\
    C 13    & 1     & 0     & 0     & G6 & 29.50 & 5.76  & 0.27  & 0.589 & 1.070 & 1.295 & 0.22  & 1.70 & 0.29\\
    C 14    & 0     & 1     & 0     & G1 & 33.70 & 3.65  & 0.24  & 0.587 & 1.030 & 1.174 & 0.23  & 1.43 & 0.81\\
    C 15    & 0     & 1     & 0     & G2 & 34.75 & 1.94  & 0.22  & 0.581 & 1.020 & 1.183 & 0.21  & 1.55 & 0.49\\
    C 16    & 0     & 1     & 0     & G3 & 35.21 & 1.45  & 0.25  & 0.586 & 1.010 & 1.141 & 0.21  & 1.35 & 0.62\\
    C 17    & 0     & 1     & 0     & G4 & 33.14 & 3.22  & 0.27  & 0.593 & 1.040 & 1.144 & 0.25  & 1.38 & 0.97\\
    C 18    & 0     & 1     & 0     & G5 & 32.79 & 4.86  & 0.23  & 0.593 & 1.040 & 1.175 & 0.26  & 1.41 & 0.61\\
    C 19    & 0     & 1     & 0     & G6 & 33.21 & 3.45  & 0.25  & 0.571 & 1.020 & 1.183 & 0.26  & 1.36 & 0.67\\
    C 20    & 0     & 0     & 1     & G1 & 37.63 & 2.25  & 0.30  & 0.625 & 1.060 & 1.177 & 0.29  & 1.52 & 1.05\\
    C 21    & 0     & 0     & 1     & G2 & 38.06 & 3.61  & 0.29  & 0.605 & 1.080 & 1.145 & 0.21  & 1.26 & 2.67\\
    C 22    & 0     & 0     & 1     & G3 & 38.96 & 2.87  & 0.30   & 0.583 & 1.080 & 1.210 & 0.22  & 1.26 & 0.75\\
    C 23    & 0     & 0     & 1     & G4 & 38.15 & 4.08  & 0.33  & 0.610 & 1.070 & 1.220 & 0.24  & 1.39 & 0.61\\
    C 24    & 0     & 0     & 1     & G5 & 37.02 & 2.35  & 0.34  & 0.609 & 1.050 & 1.260 & 0.26  & 1.78 & 0.82\\
    C 25    & 0     & 0     & 1     & G6 & 37.38 & 3.89  & 0.30   & 0.615 & 1.060 & 1.240 & 0.25  & 1.83 & 0.59\\
    \midrule
    \end{tabular}%
  \label{tab:PARAMETI_}%
\end{table*}%

All these sets of parameters lead to comparable cost functions and can thus be considered `optimal' candidates if evaluated in the data used for their derivation. However, as we shall see shortly, their fitness decreases significantly when evaluated on different data. 
Moreover, some parameters are more variable than others. This is further illustrated in Table \ref{tab:varianza-par}, which shows the average value for each parameter, theirs standard deviation to mean ratio, and theirs minimum and maximum value. It is thus clear that the parameter $h_s, \alpha, \beta$ vary significantly among the various calibration, contrary to $e_{i0}, e_{d0}, e_{c0}$.

\begin{table}[tb]
 \centering
  \caption{Statistics of the obtained parameters. Mean ($\mu$), standard deviation ($\sigma$) over mean, minimum and maximum. }
    \begin{tabular}{llrrrr}
    \toprule 
    \multicolumn{2}{l}{Par.} & \multicolumn{1}{c}{$\mu$ } & \multicolumn{1}{c}{$\sigma/\mu\cdot 10^2$} & \multicolumn{1}{c}{$min$} & \multicolumn{1}{c}{ $max$}\\
    \midrule
    $\varphi_c$&($^{\circ}$) &  33.65 & 9.7 & 25.32 & 38.96  \\
    $h_s$    &($GPa$)      &  4.18 &  42.7 & 1.45 &  7.86 \\
    $n$      &($-$)        &  0.28 &  11.0 & 0.22 & 0.34  \\
    $e_{d0}$ &($-$)        &  0.60 &  4.9 & 0.55  & 0.67  \\
    $e_{c0}$ &($-$)        &  1.04 &  3.9 & 0.94  & 1.08  \\
    $e_{i0}$ &($-$)        &  1.20 &  3.7 & 1.14  & 1.30  \\
    $\alpha$ &($-$)        &  0.24 &  20.1 & 0.10  & 0.29 \\
    $\beta$  &($-$)        &  1.53 &  18.6 & 1.17  & 2.50  \\
    \bottomrule
    \end{tabular}%
  \label{tab:varianza-par}
\end{table}%

To evaluate the accuracy of each set of parameters, we now consider the mean deviation in the abscissa of each plane ($\sigma_v,\,e$), ($\varepsilon_a,\,q$) and ($\varepsilon_a,\,\varepsilon_v$), normalised by the corresponding maximum variation in the experimental data from each test. That is, given 

\begin{equation}
    \xi_i=\frac{|Y_{num}-Y_{dat}|}{max(Y_{dat})-min(Y_{dat})}\,\cdot 100\,,
\end{equation} with $Y_{num}$ the SH model prediction and $Y_{dat}$ the experimentally measured ones, in each plane, we define the global deviation as 

\begin{equation}
\label{xi}
    \xi=\mathbb{E}(\xi_i),
\end{equation} where, $\mathbb{E}$ is the expectation operator. 

Although this fitness measurement is less suited than the Fréchet distance for calibration purposes because it evaluates the model errors only in one of the axes in each plane, it allows for a more intuitive evaluation of the model performances in relation to the data variability. Indicatively,  $\xi\leq 20 \%$ is acceptable for the CD test and $\xi\leq 30 \%$ for the OE test.

\begin{figure*}[h]
  \centering
    \includegraphics[width=0.93\textwidth]{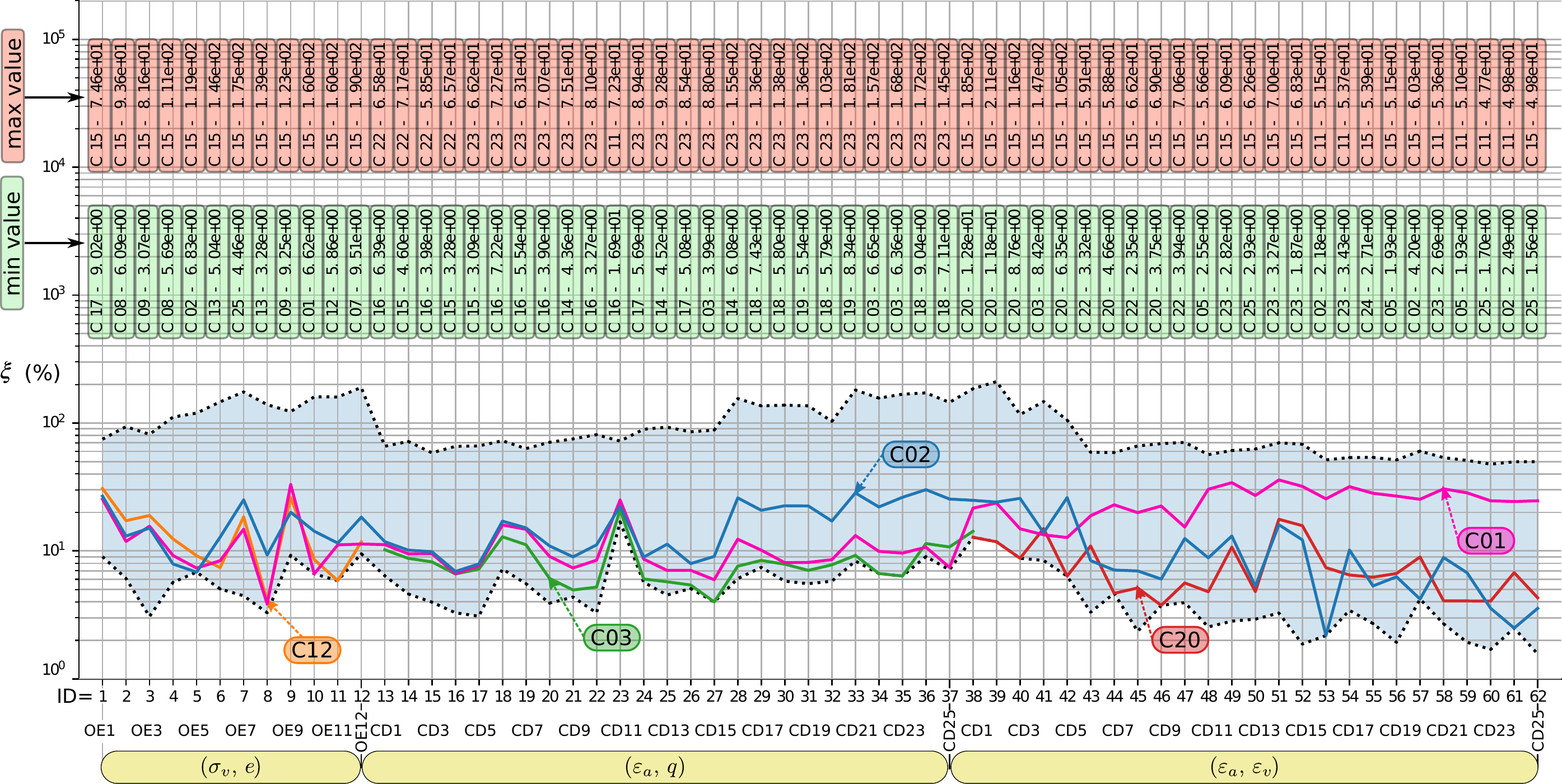}
\caption{Average percentage deviation $\xi$ \eqref{xi} in all the OE and CD test data (Tables \ref{tab:OED_init} and \ref{tab:TxD_in}) considered in this work. The abscissa recalls the test ID and the plane at which the deviation is measured. The combinations providing the best and worst performances for each test are recalled at the top. In the graph, the dashed area in light blue (delimited by black dashed lines) is the envelope of performances for all the combinations of parameters in Table \ref{tab:PARAMETI_}). The performance of some illustrative combinations is also reported. The blue, orange, green and red lines are associated with the combinations C15, C12, C03, C20, respectively. These combinations excel in some planes (e.g. C15 and C20 in the second plane, C12 in the first plane, etc.) but fail in others.}
\label{fig:COST_Lin} 
\end{figure*}

The SH model performances using all the sets in Table \ref{tab:PARAMETI_} is illustrated in detail in Figure \ref{fig:COST_Lin}. The $x$ axis provides the ID of each test and the plane in which the deviation $\xi$ is measured. The first 12 IDs are given to the OE tests, evaluated in the $(\sigma_v, \, e)$ plane. The IDs from 12 to 37 are given to the CD tests evaluation in the ($\varepsilon_a ,\,q$) plane, while the last IDs, from 37 to 62, are given to the CD tests evaluated in the plane ($\varepsilon_a,\,\varepsilon_v$).  

For each test, the upper portion of the graph indicates the minimum and the maximum deviation together with the associated parameter combination. The bottom portion of the graph shows the envelope of performances for all combinations, together with some illustrative cases. No combination reaches the minimal error in all planes. For the OE tests, $\xi$ varies from $3.0$ to $189.6$ \%, for the CD it varies from $3.1$ to $181.1$ \%, in the $ (\varepsilon_a , \, q)$ plane and from $1.6$ to $210.9$ \%, in the $(\varepsilon_a , \, \varepsilon_v)$ plane. 

\begin{figure}[tb]
  \centering
  \subcaptionbox{Oedometric plane}[1\linewidth][c]{%
    \includegraphics[width=0.44\textwidth]{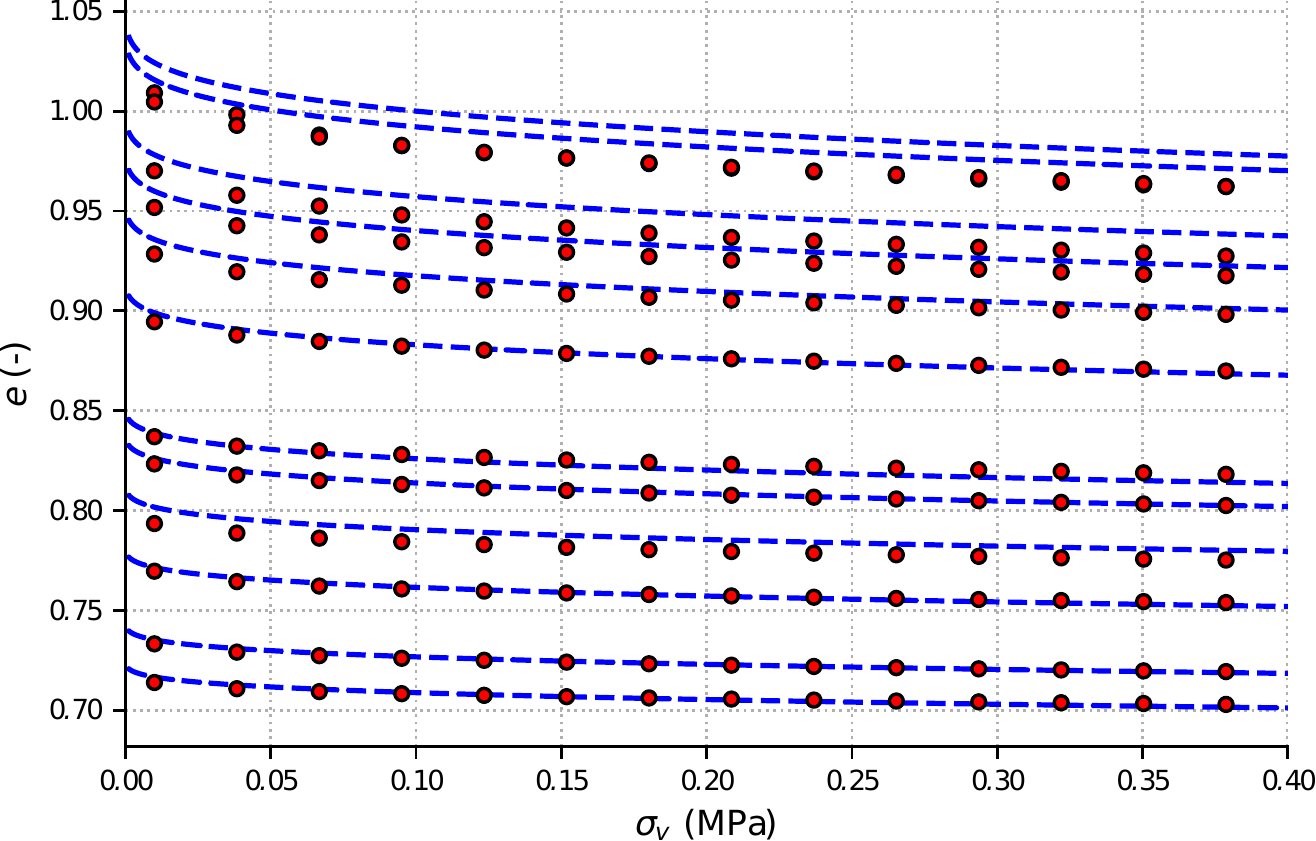}}\\
  \subcaptionbox{Triaxial deviatoric plane}[1\linewidth][c]{%
    \includegraphics[width=0.44\textwidth]{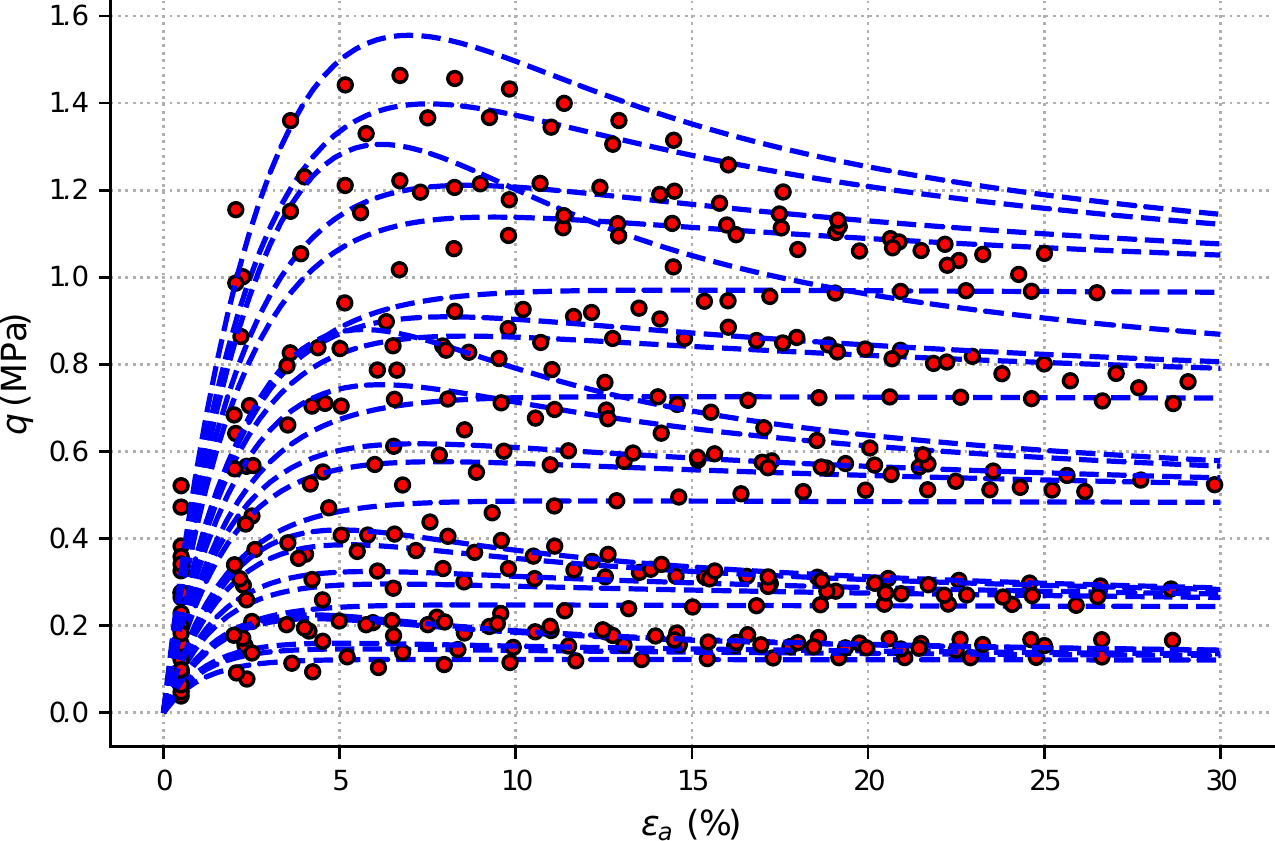}}\\
  \subcaptionbox{Triaxial volumetric plane}[1\linewidth][c]{%
    \includegraphics[width=0.44\textwidth]{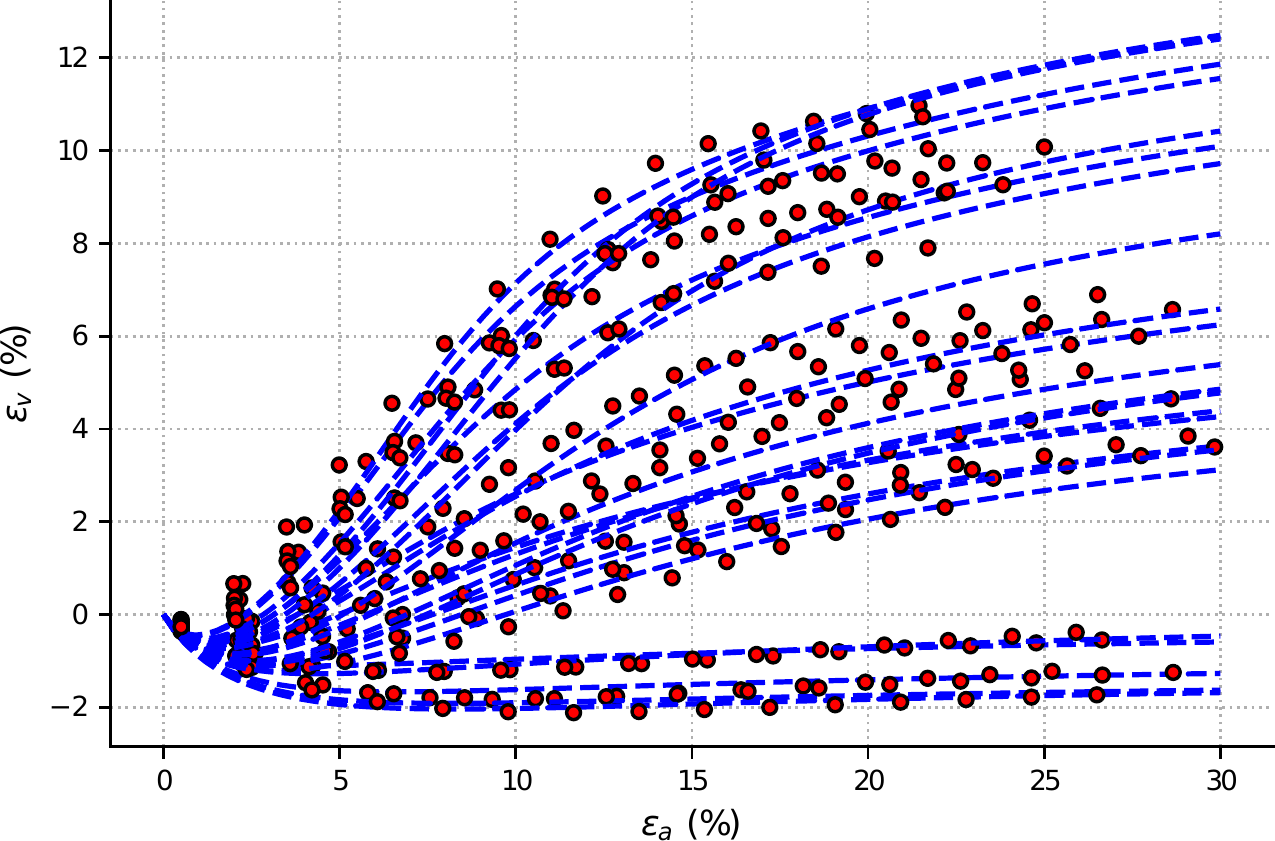}}
  \caption{Calibration of the SH model for the individual planes. The figures display the curves obtained with the combination C12, C03 and C20 in the three different planes (a),(b), and (c). These three combinations accurately interpret the results for the 12 OE and 25 CD tests.}
\label{fig:GA12GA3G6}
\end{figure}

For example, the combination C12 provides excellent results in the CD in the ($\sigma_v,\, e$), with $\xi\in [4,30.7]\%$  and mean $14.5\%$, but fails in both ($\varepsilon_a,\, q$) and ($\varepsilon_a,\,\varepsilon_v$) planes. Not surprisingly, the GA obtained this combination when the weights were set to $(1,0,0)$. The trends for the combinations C03 and C20 are also shown using green and red lines in Figure \ref{fig:COST_Lin}. The combination C20 performs well in the plane ($\varepsilon_a, \,\varepsilon_v$) with an $\xi\in [3.7,17.7\%]$ and mean $7.9\%$, while the combination C03 perform well in the plane ($\varepsilon_a ,\, q$) and $\xi\in [3.9,21.0\%]$ and mean $8.5\%$. While this is in line with the choice of weights for C20 and C12, the combination C03 gave equal weights to all planes, thus showing that the amount and the kind of data used for the training play an important role. Figure \ref{fig:GA12GA3G6} provides a qualitative overview of the good agreement between the experimental data and the numerical prediction with the combinations C12, C03 and C20. 

\begin{figure}[tb]
  \centering
  \subcaptionbox{Oedometric plane}[1\linewidth][c]{%
    \includegraphics[width=0.44\textwidth]{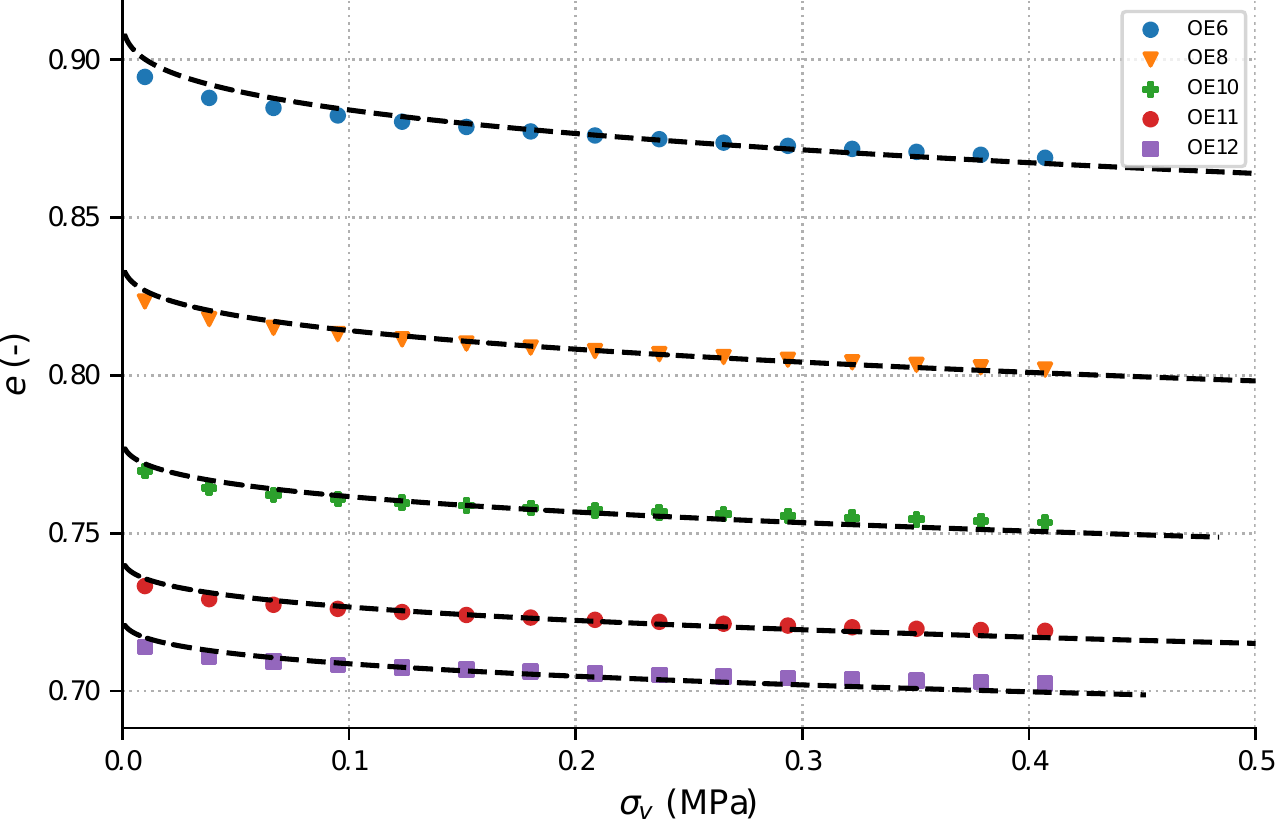}}\\
  \subcaptionbox{Triaxial deviatoric plane}[1\linewidth][c]{%
    \includegraphics[width=0.44\textwidth]{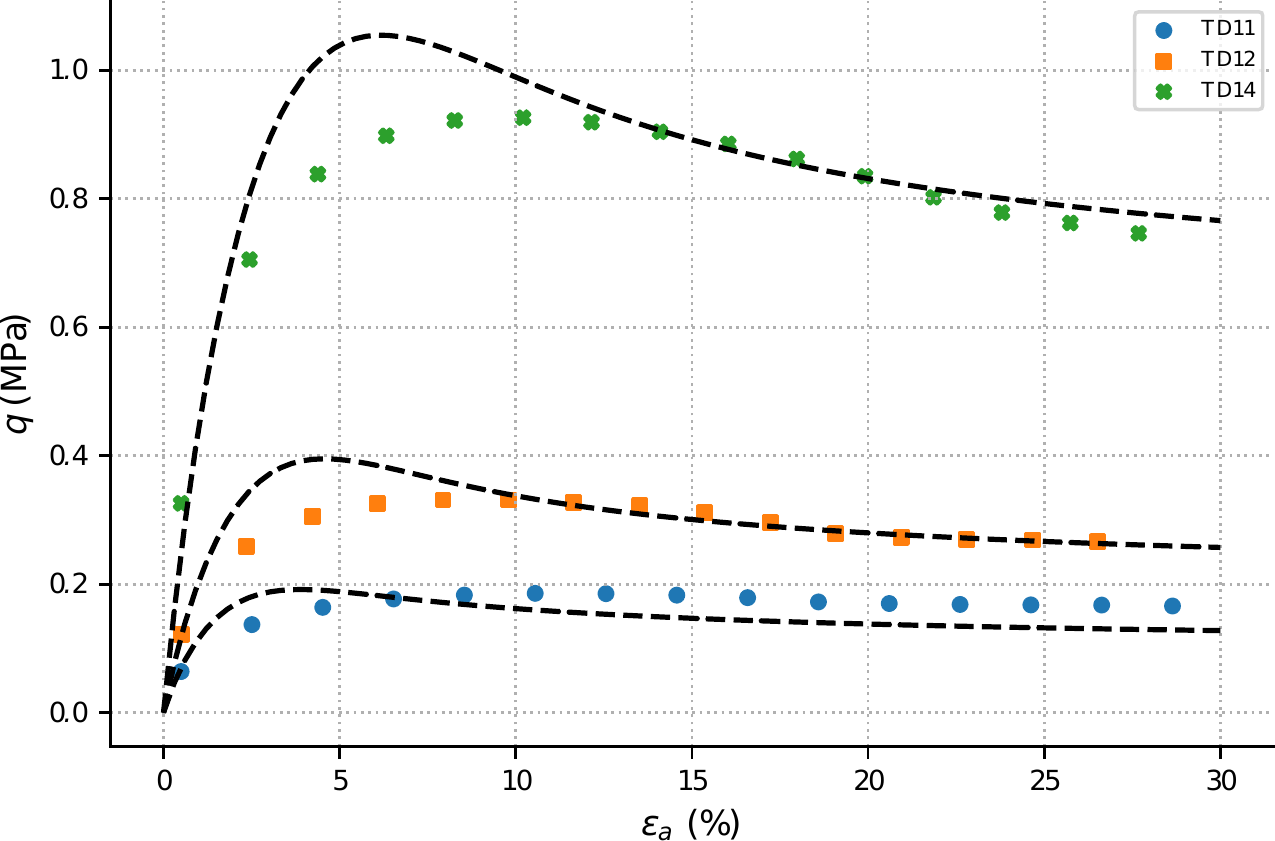}}\\
  \subcaptionbox{Triaxial volumetric plane}[1\linewidth][c]{%
    \includegraphics[width=0.44\textwidth]{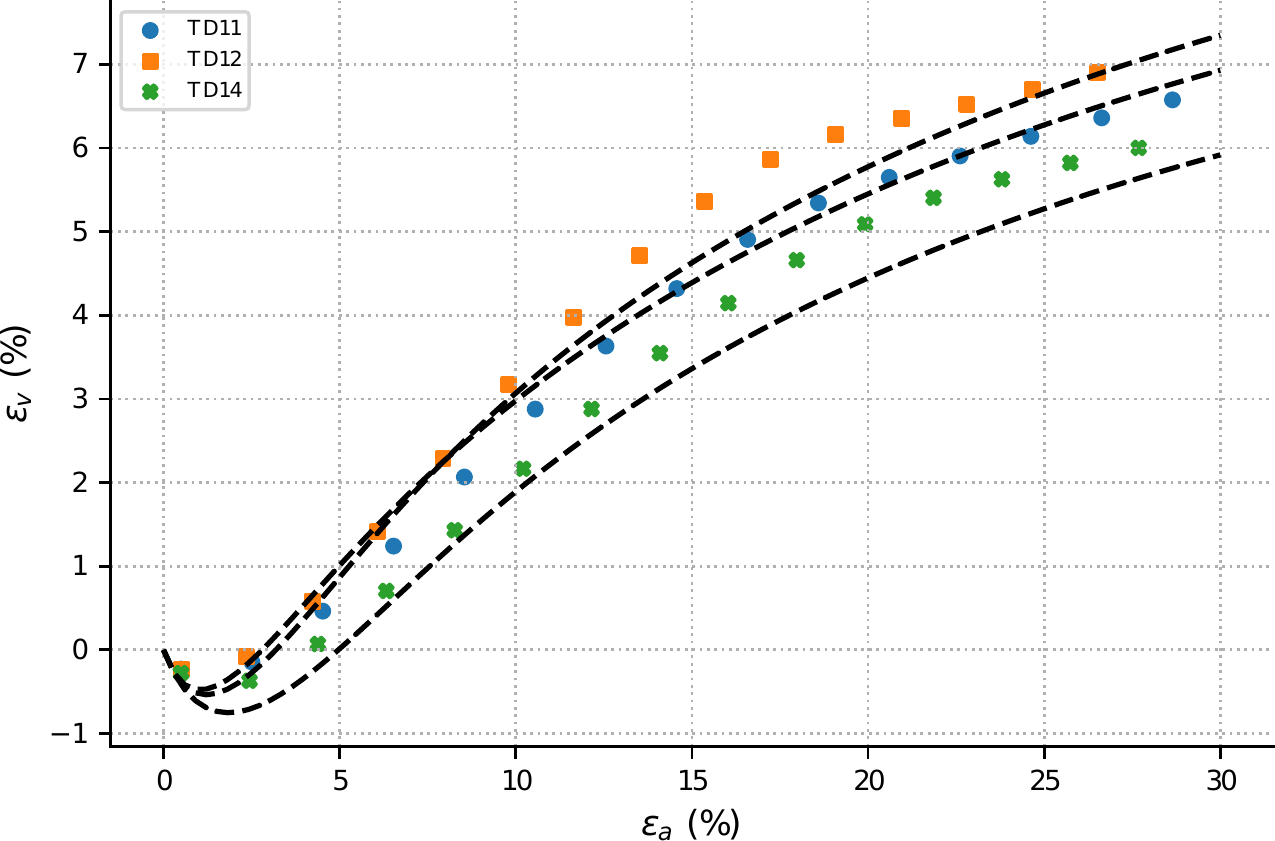}}
  \caption{Performance of the SH model in the simultaneous description of CD and OE tests. The figures display the curves obtained with the combination C05. With this combination, we get a good result in all three planes simultaneously for the (CD11, CD12, and CD13).}
\label{fig:GA5}
\end{figure}

As shown in Figure \ref{fig:GA5}, some experiments (CD11, CD12, CD13) can be well described with a single set of parameters ($C05$). We have $\xi\in[2.5,23.4]\%$, with a mean of $ 9.1 \% $, for these tests.

Figure \ref{fig:COST_Lin} also shows the average percentage deviation for the SH using the model parameters by Wichtman and Triantafyllidis (WT) \cite{Wichtmann2016-rs} (combination C01) and for the combination C02, which was derived using all the dataset and giving all planes equal importance. Both C01 and C02 perform worst than the optimal combinations in each test, and the comparison shows that C01 favours the model accuracy in the planes ($\sigma_v,\, e$) and ($\varepsilon_a,q$) over the ones in the plane ($\varepsilon_a, \,\varepsilon_v$).

\begin{figure*}[htb]
  \centering
  \subcaptionbox{Oedometric plane}[0.45\linewidth][c]{%
    \includegraphics[width=0.428\textwidth]{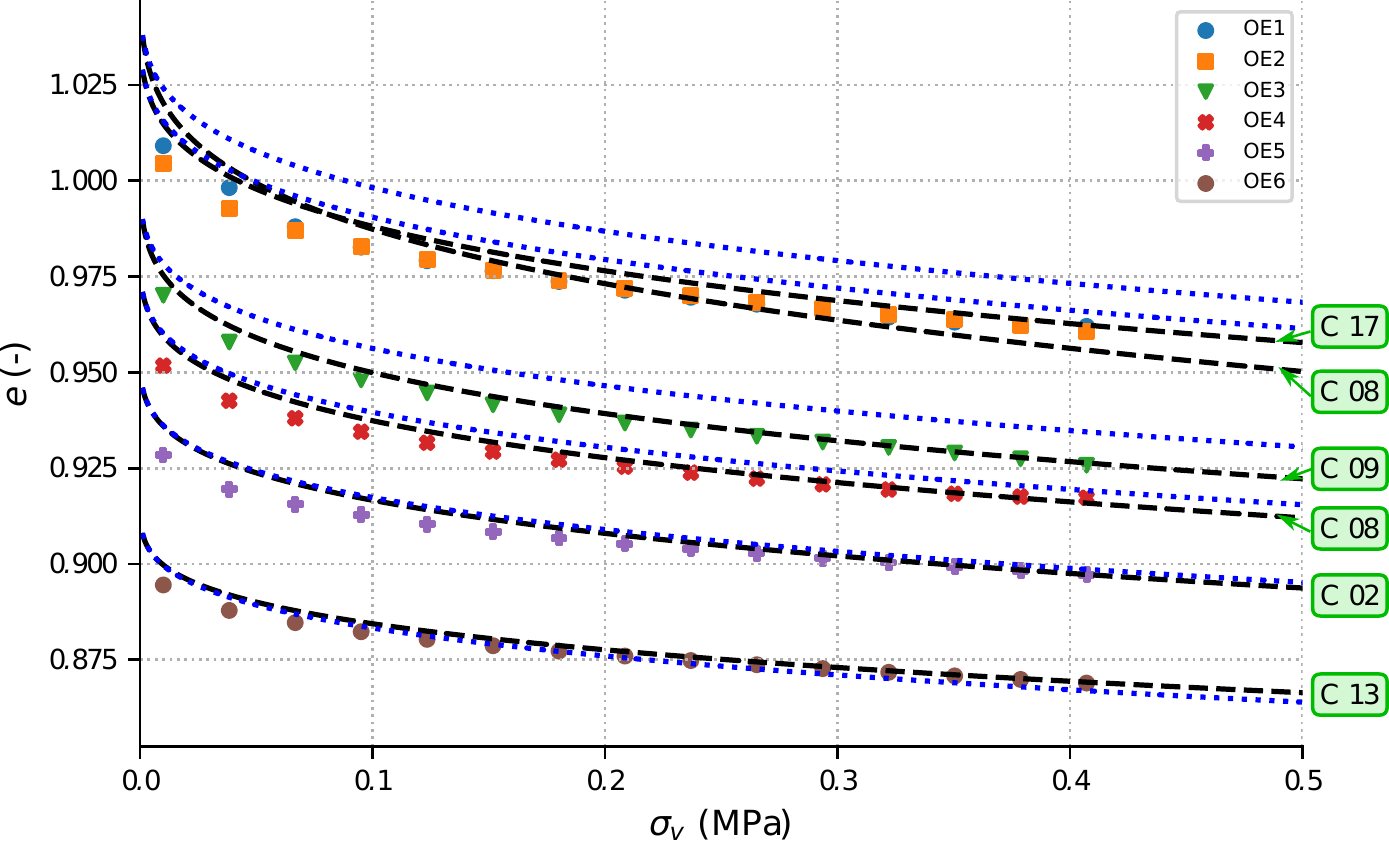}}\hfil
  \subcaptionbox{Oedometric plane}[0.45\linewidth][c]{%
    \includegraphics[width=0.428\textwidth]{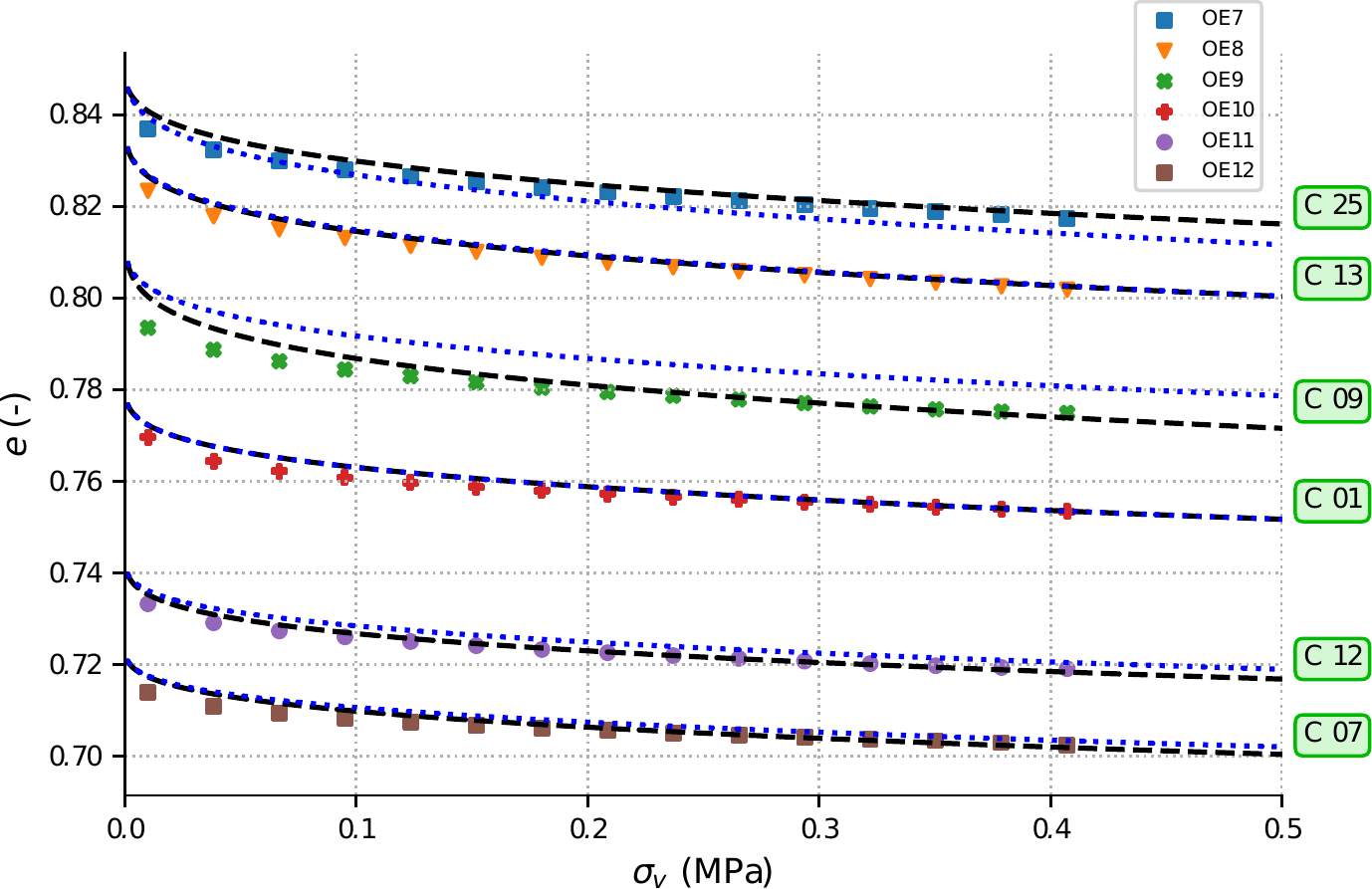}}\\
  \subcaptionbox{Triaxial deviatoric plane}[0.45\linewidth][c]{%
    \includegraphics[width=0.416\textwidth]{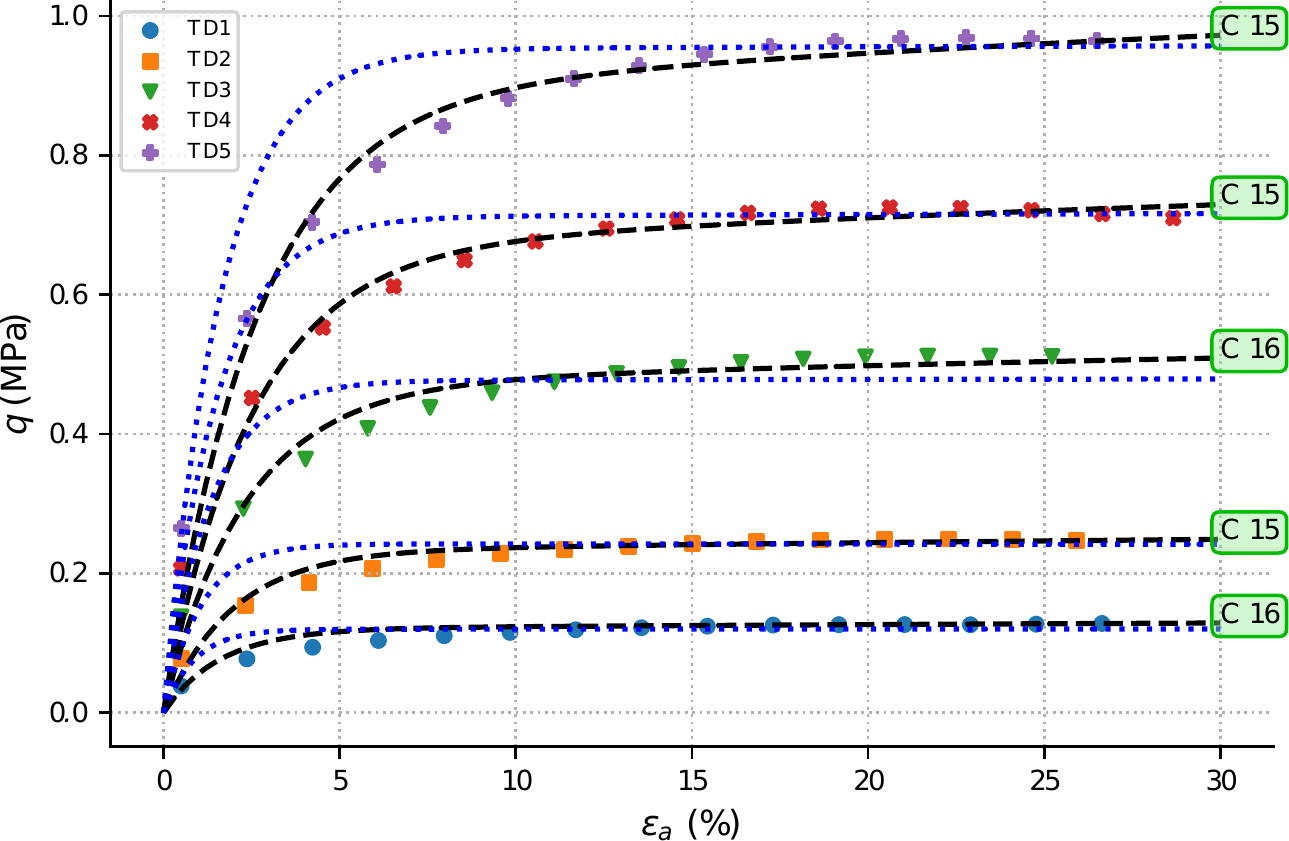}}\hfil
    \subcaptionbox{Triaxial deviatoric plane}[0.45\linewidth][c]{%
    \includegraphics[width=0.416\textwidth]{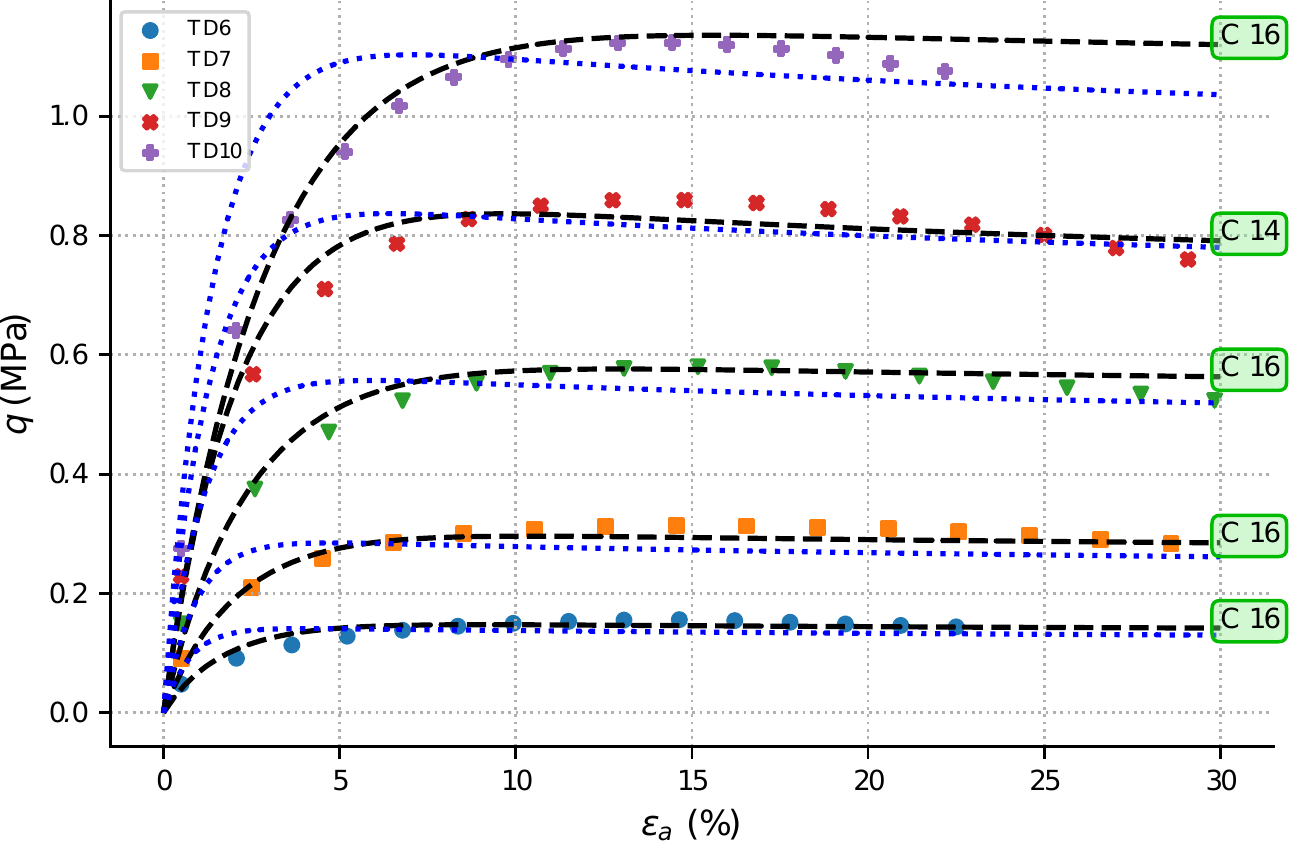}}\\
   \subcaptionbox{Triaxial deviatoric plane}[0.45\linewidth][c]{%
    \includegraphics[width=0.416\textwidth]{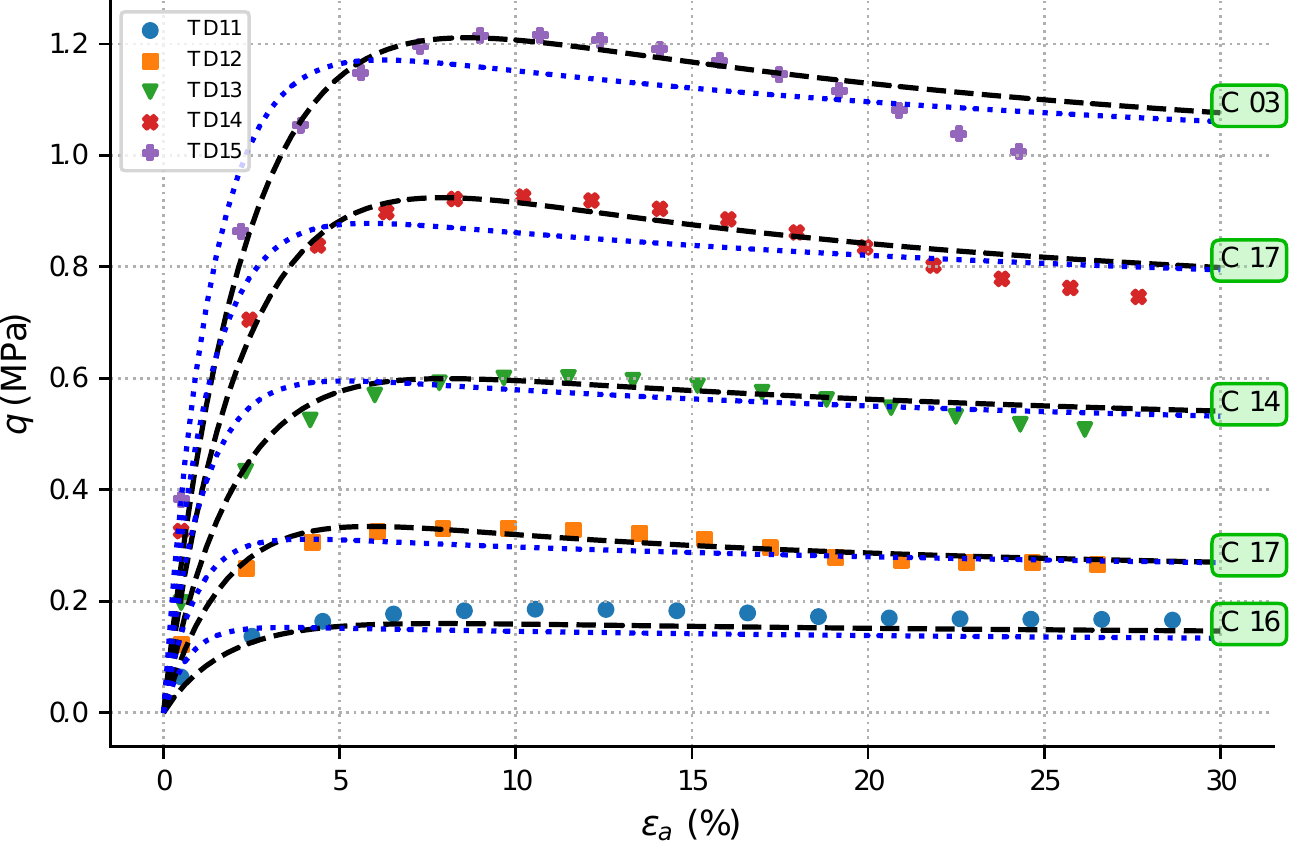}}\hfil
    \subcaptionbox{Triaxial deviatoric plane}[0.45\linewidth][c]{%
    \includegraphics[width=0.416\textwidth]{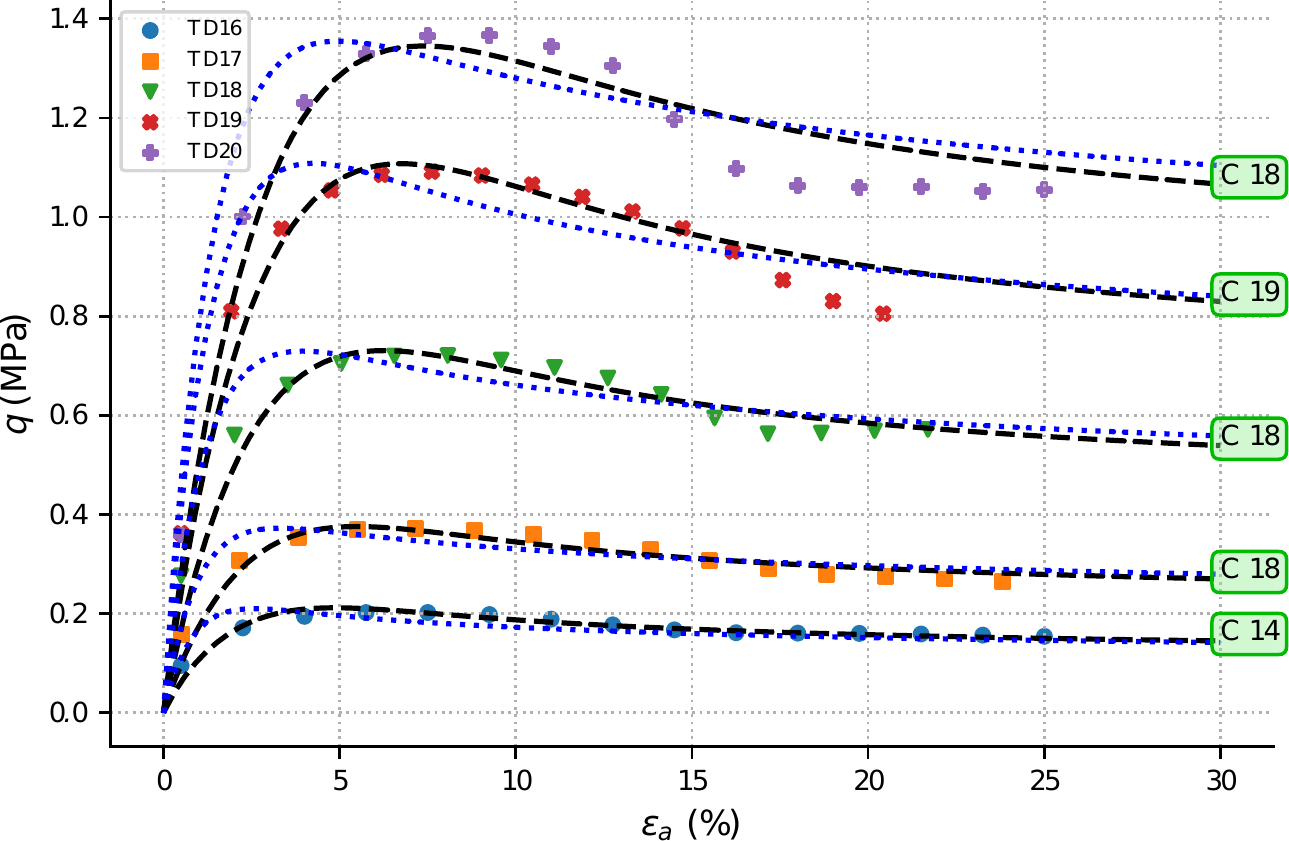}}\\
   \subcaptionbox{Triaxial deviatoric plane}[0.45\linewidth][c]{%
    \includegraphics[width=0.416\textwidth]{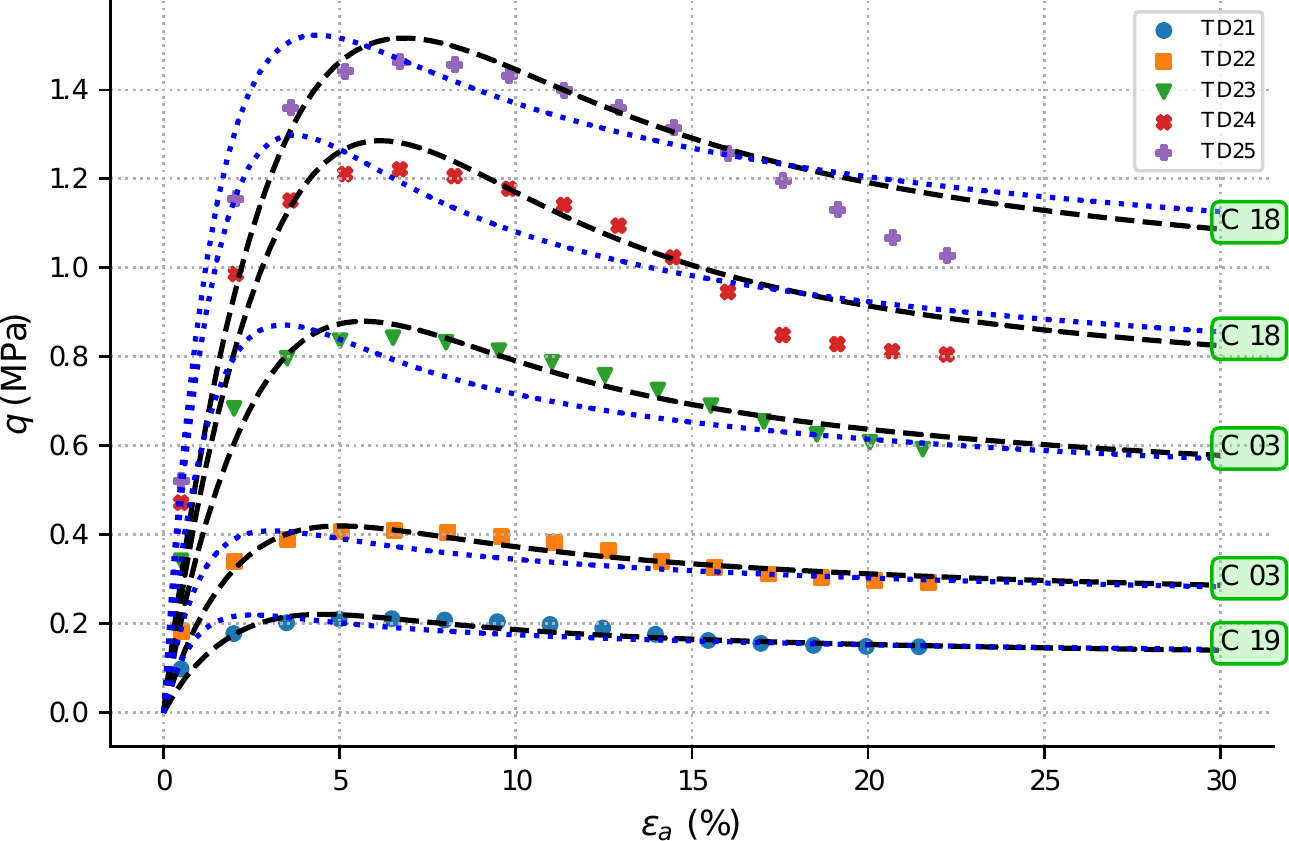}}\hfil
    \subcaptionbox{Triaxial volumetric plane}[0.45\linewidth][c]{%
    \includegraphics[width=0.40\textwidth]{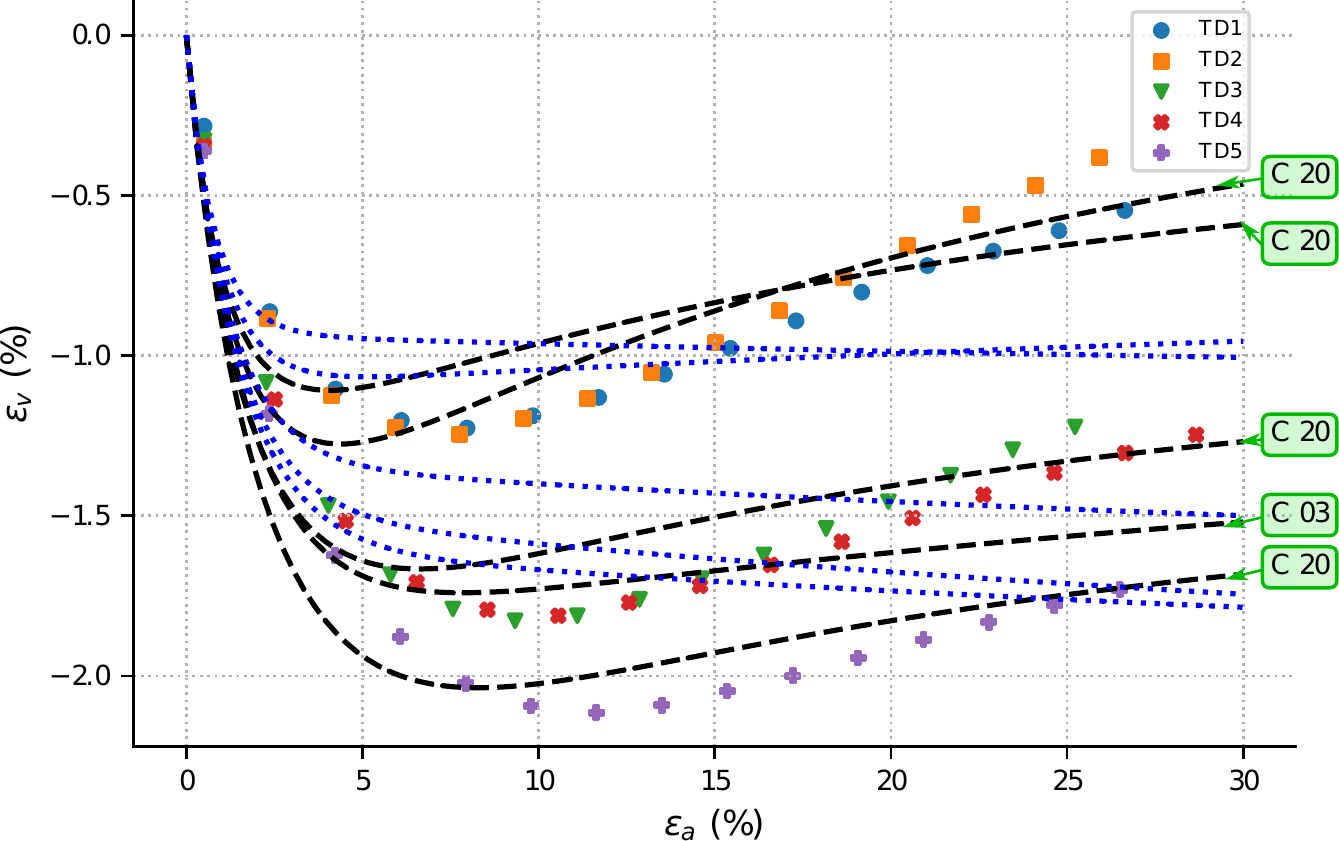}}\\
  \caption{Comparison between the response curves of the SH model and the experimental data (point). The blue dotted lines are computed using the parameters by Wichtman and Triantafyllidis (C01), and the black dashed line with the parameters obtained with GA-cal. The label of the curve refers to Table \ref{tab:PARAMETI_} - Part I.}
\label{fig:GA_best_P1}
\end{figure*}

\begin{figure*}[htb]
  \centering
  \subcaptionbox{Triaxial volumetric plane}[0.45\linewidth][c]{%
    \includegraphics[width=0.40\textwidth]{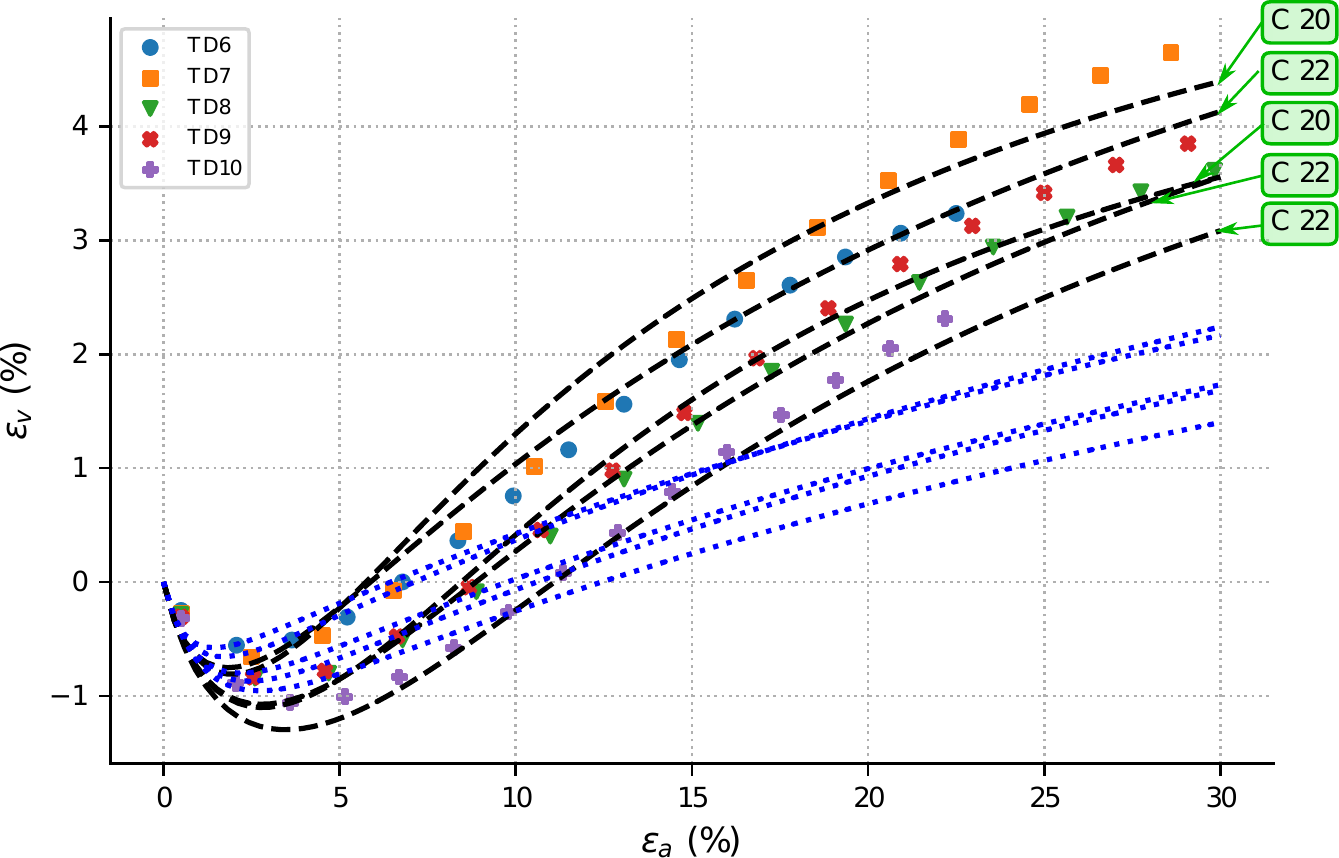}}\hfil
  \subcaptionbox{Triaxial volumetric plane}[0.45\linewidth][c]{%
    \includegraphics[width=0.40\textwidth]{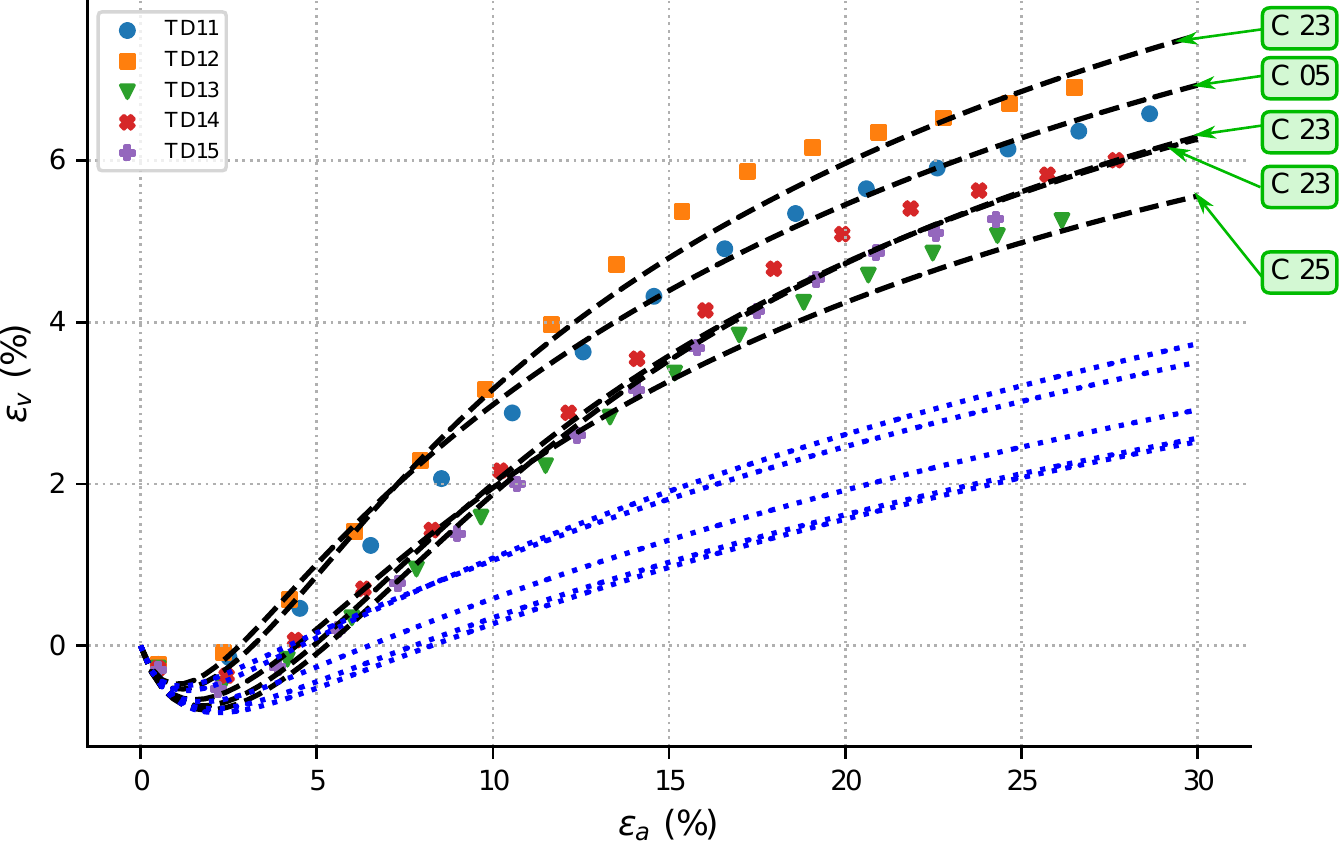}}\\
  \subcaptionbox{Triaxial volumetric plane}[0.45\linewidth][c]{%
    \includegraphics[width=0.40\textwidth]{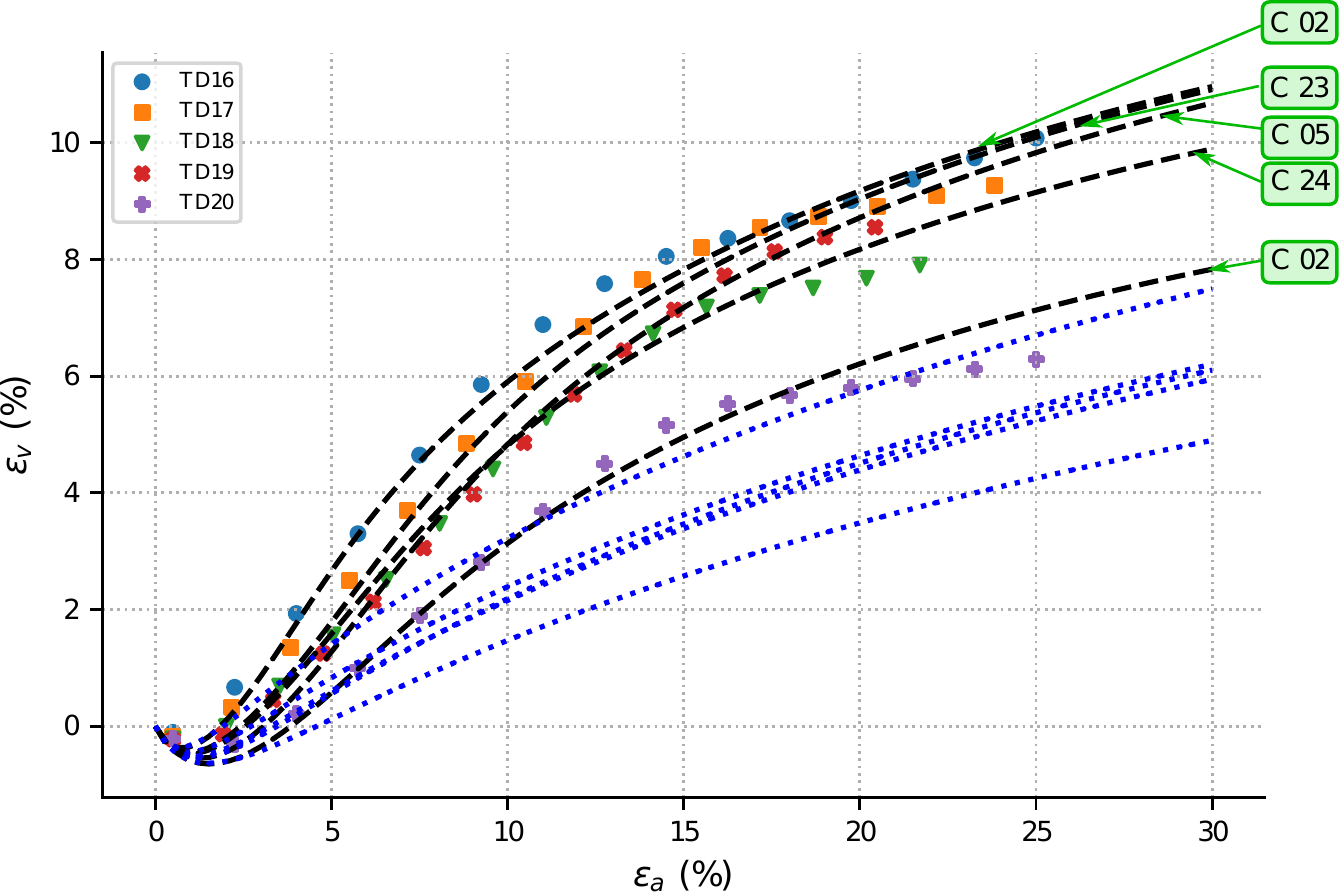}}\hfil
    \subcaptionbox{Triaxial volumetric plane}[0.45\linewidth][c]{%
    \includegraphics[width=0.40\textwidth]{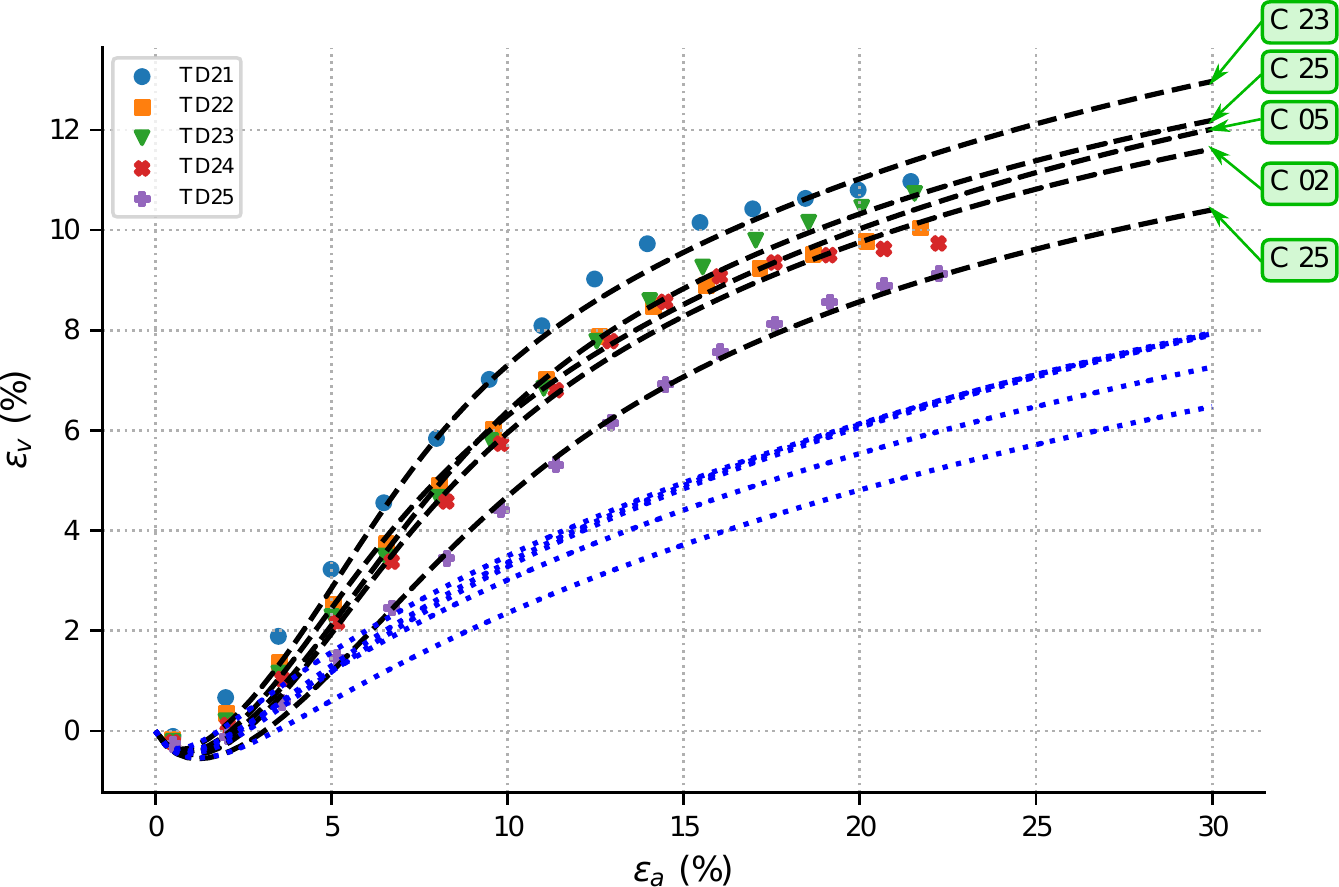}}
  \caption{Comparison between the response curves of the SH model and the experimental data (point). The blu dotted lines are computed using Wichtman and Triantafyllidis (C01) parameters, and the black dashed line with the parameters obtained with GA-cal. The label of the curve refers to Table \ref{tab:PARAMETI_} - Part II.}
\label{fig:GA_best_P2}
\end{figure*}

To further illustrate the performances of the different combinations, Figures \ref{fig:GA_best_P1} and \ref{fig:GA_best_P2} plot the experimental data 
together with the model predictions using the best combination on that specific set (black dashed lines) and the predictions using the parameters by Wichtman and Triantafyllidis (C01, with the blue dotted line). While all the identified optimal combinations outperform C01, the improvement comes at the cost of considering a different set of parameters for each test.

The relative importance of model accuracy in these planes depends significantly on the application. For example, in the slope stability analysis, one is more interested in the accurate modelling of the soil response in the plane ($\varepsilon_a,q$), and the combination C03 could suit that purpose. On the other hand, in the settlement prediction of a shallow foundation, one is primarily interested in the soil response in the plane ($\varepsilon_a,\,\varepsilon_v$), and hence combination C06 could be more appropriate.

In problems such as soil-structure interaction, where the behaviour in all planes are equally relevant, one must accept a compromise (see combination C02) and use a set of coefficients that is sub-optimal in each plane taken separately. The main conclusion in this analysis is that the model calibration in these three planes yields contrasting objectives that bring the SH model to its limits. We search for the parameters driving the compromise in the following section.

\subsection{The Pareto Front} \label{R2}
The analysis of the combination C01 and C02, aimed at modelling the soil response in all planes, shows that the main discrepancies occurs in the planes $(\varepsilon_a,q)$ and $(\varepsilon_a,\varepsilon_v)$, in that the success in one of them yields failure in the other.

\begin{figure}[tb]
  \centering
  \subcaptionbox{Pareto front ($\delta_2,\delta_3$) coloured with  $\alpha$ \label{fig:Pareto_a}}[1\linewidth][c]{%
    \includegraphics[width=0.49\textwidth]{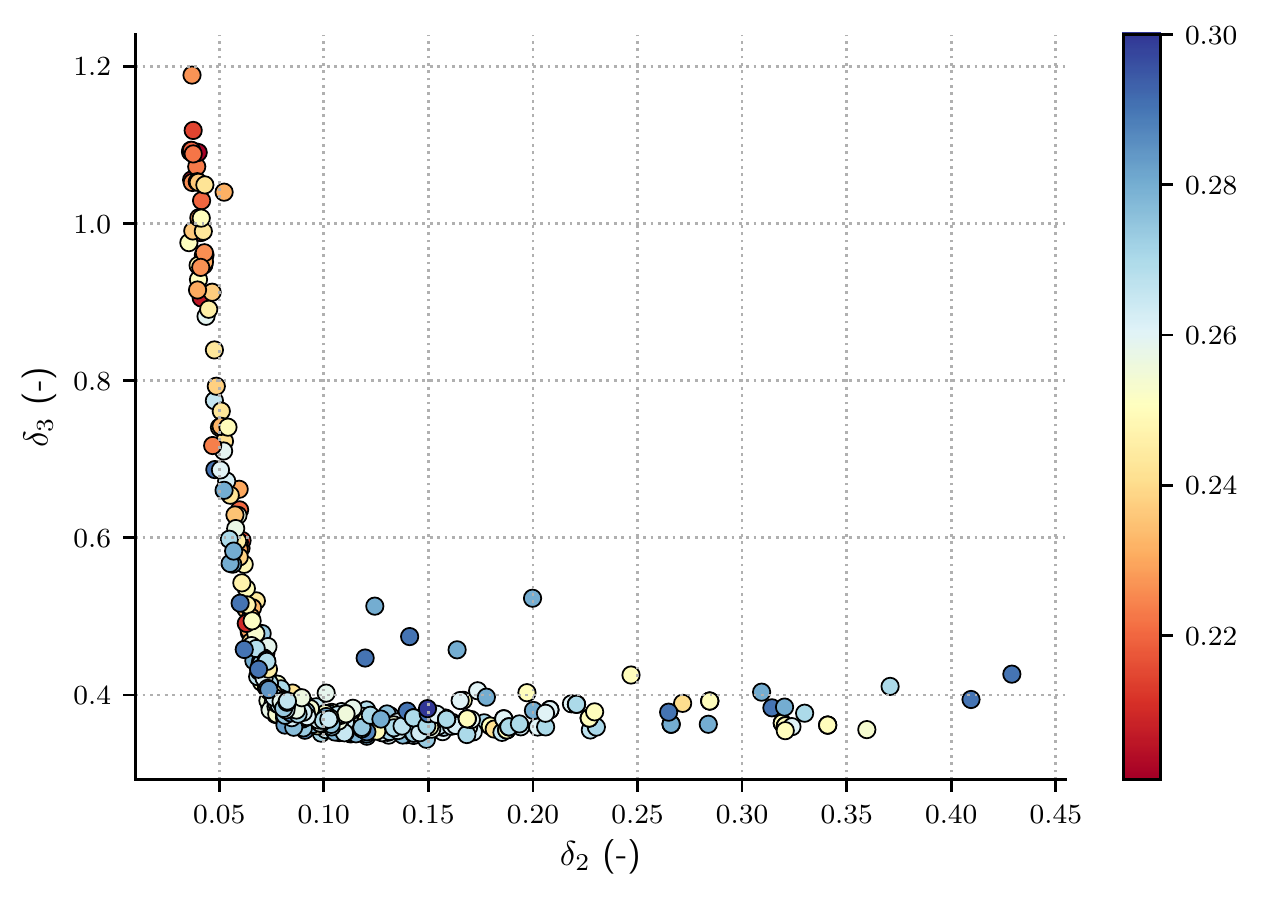}}\\
  \subcaptionbox{Pareto front ($\delta_2,\delta_3$) coloured with  $\varphi_c$.}[1\linewidth][c]{%
    \includegraphics[width=0.49\textwidth]{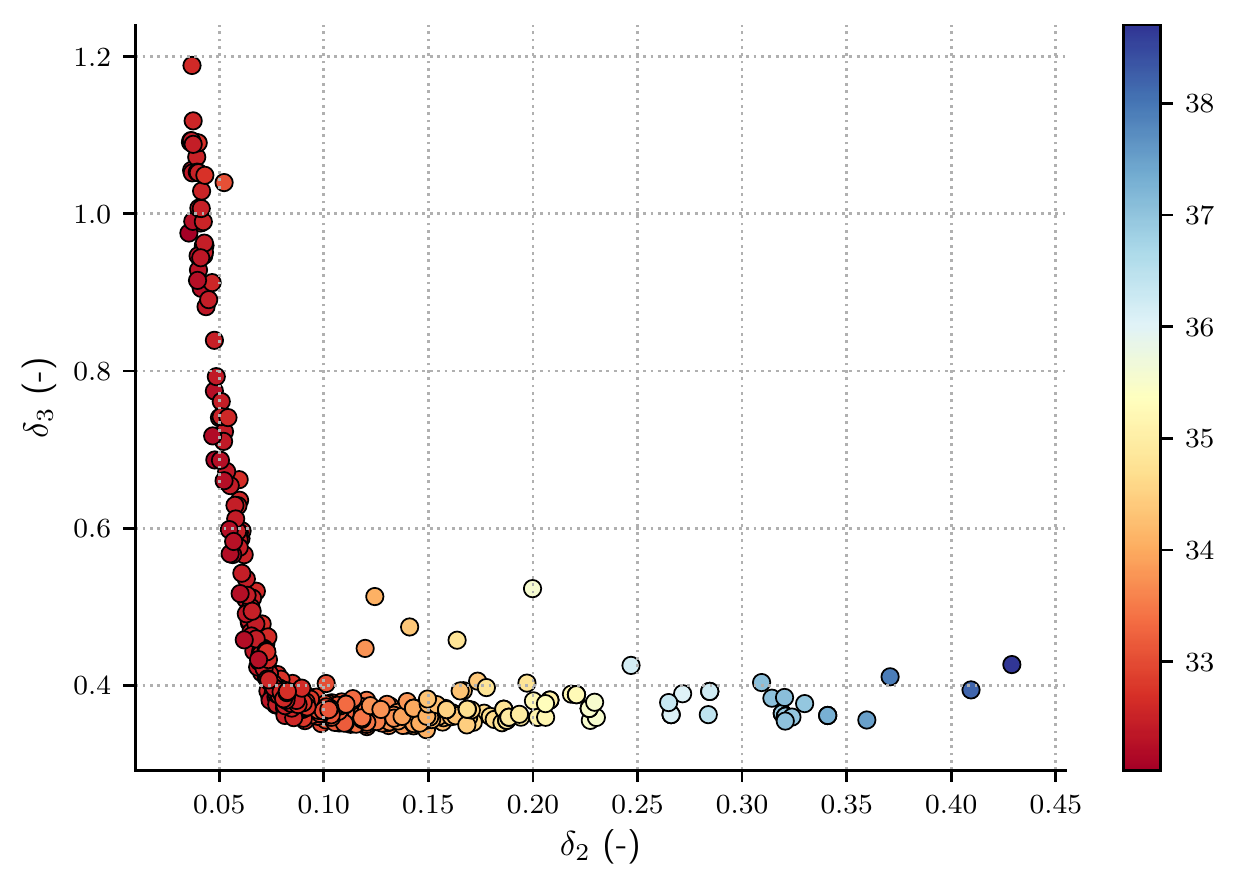}}
  \caption{Search for the Pareto front in the minimisation of $\delta_2$ (model performances in the ($\varepsilon_a,\,q$) plane) and $\delta_3$ (model performances in the ($\varepsilon_a,\,\varepsilon_v$) plane). The marker's location is computed by averaging the errors on the 25 tests. These are coloured with the calibrated value for  $\alpha$ (a), $\varphi_c$ (b) and $e_{d0}$ (c).}
\label{fig:Pareto}
\end{figure}

In this section, we further analyse the performances in these two planes and identify which parameters are mostly involved in the conflicting requirements. Specifically, we treat the calibration as a multi-objective optimisation which seeks to minimise the average Fréchet distances in these two planes, i.e. $\delta_2$ and $\delta_3$ in \eqref{eq:cost}. We then search for the optimal trade-offs between the two objectives, i.e. the Pareto front \cite{Fonseca1995AnOO} for the problem. At the Pareto front, improvements in one plane cannot be attained without worsening in the other. 

The search for the Pareto front was conducted using the weighted sum method \cite{Kim2006}. This approach consists in performing a large set of single objective optimisations (in this case, 400) using weights defined as 

\begin{equation}
    (1,\, w_p,\, 1) \quad \text{and}\quad (1,\, 1,\, w_p) 
\end{equation} where $w_p$ is a vector of linearly spaced values from 0 to 10 with 200 elements. For each calibration, we monitor the values of $\delta_2$ and $\delta_3$, averaged over the 25 tests in Table \ref{tab:TxD_in}. The results are plotted in Figure \ref{fig:Pareto}, with markers coloured by two of the eight parameters in the optimal set obtained by the calibration: $\alpha$ and $\varphi_c$.

The Pareto front is qualitatively identified. Furthermore, the marker face colour allows us to reveal whether any of these parameters changes monotonically from one extreme of the front to the other, and is thus potentially correlated with the degradation of performances in one plane (and the consequent improvement in the other).

\begin{figure}[tb]
  \centering
  \subcaptionbox{$\nu_i$ versus the parameters $\alpha$ \label{fig:PoissonPareto_a}}[1\linewidth][c]{%
    \includegraphics[width=0.49\textwidth]{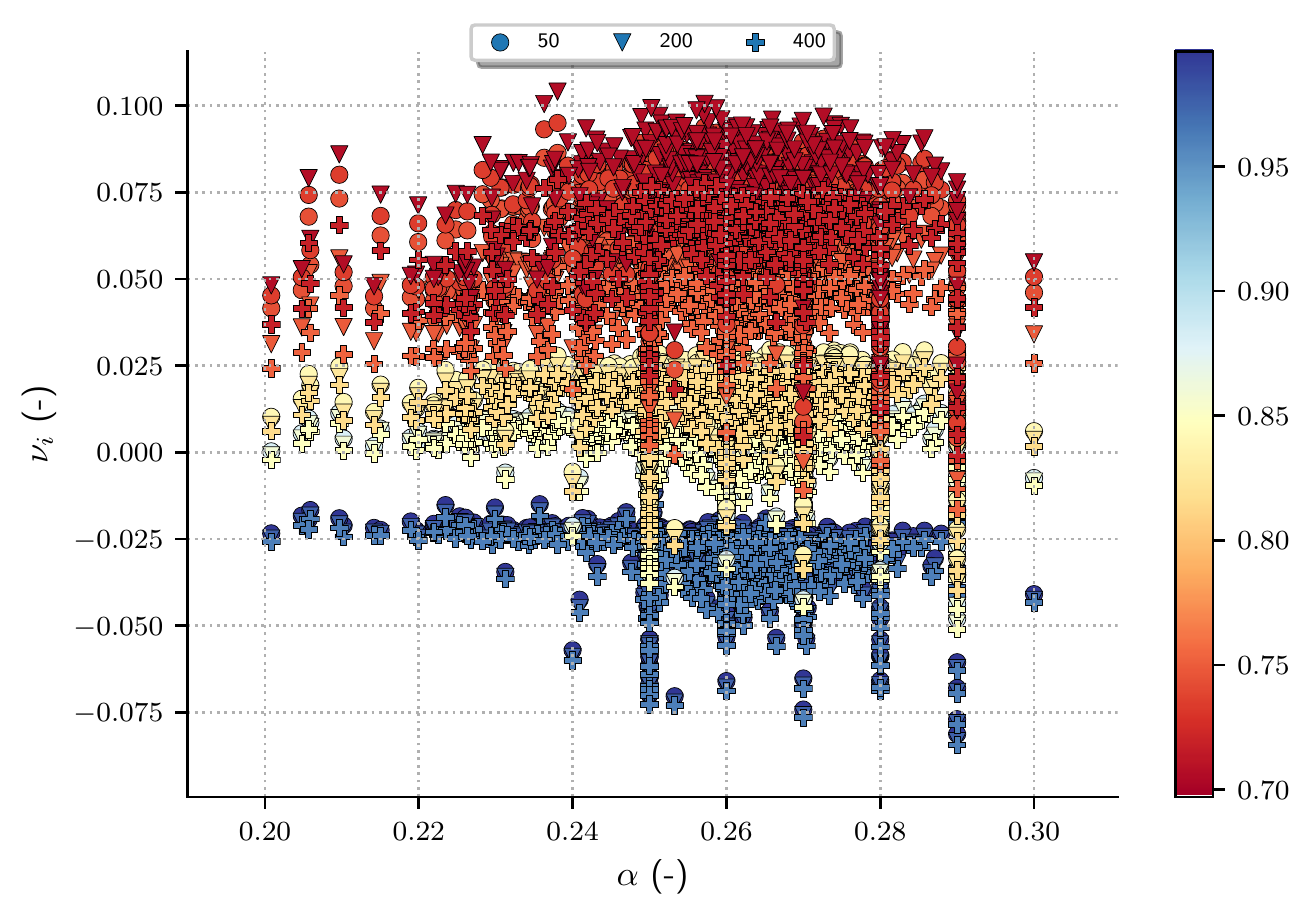}}\\
  \subcaptionbox{$\nu_i$ versus the parameters $\varphi_c$ }[1\linewidth][c]{%
    \includegraphics[width=0.49\textwidth]{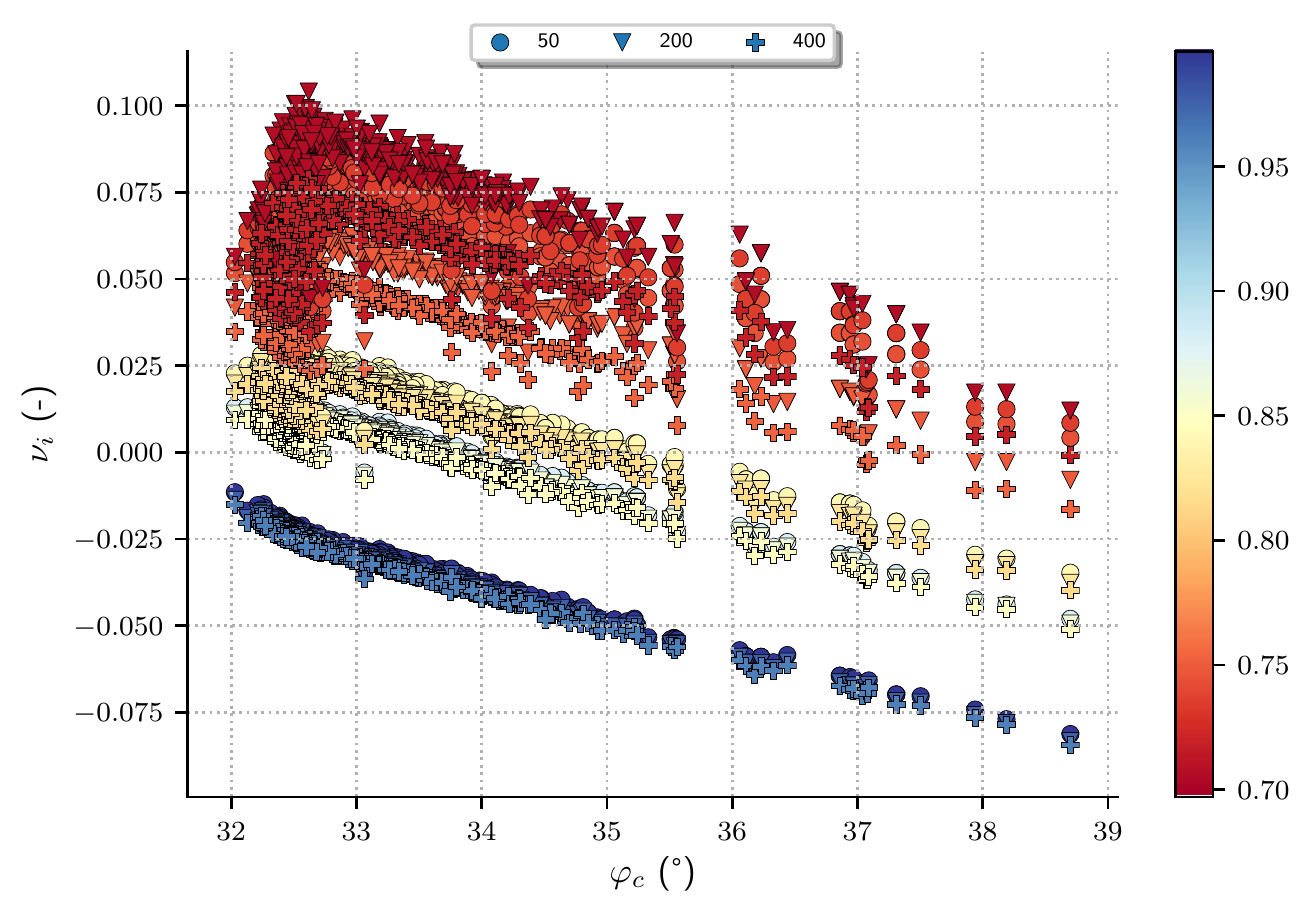}}
  \caption{Plot of the initial Poisson ratio $\nu_i$ versus the parameters $\alpha$ (a), and $\varphi_c$ (b). The figures highlight the impact of the initial conditions: different markers are used for the initial pressure $p_0$ while the colour scale in the marker's face refers to the initial void ratio $e_0$ (see table \ref{tab:TxD_in}).}
\label{fig:PoissonPareto}
\end{figure}

The parameter $\alpha$ was included because this is often considered to be the main responsible for the poor performances in the $(\varepsilon_v, q)$ plane and is usually calibrated to reproduce the peak stress rather than the volumetric deformation \cite{Wichtmann2016-xw,Kadlicek2022_II,Machacek2022}. However, the plot in Figure \ref{fig:Pareto_a} display no clear trend along the Pareto front and thus does not support this common belief. Similar observations holds for the other parameters $h_s,\,n,\,\beta,\,e_{i0},\,e_{c0}$ and $e_{d0}$, which are thus not shown.

On the other hand, $\varphi_c$ increases when moving from low $\delta_2$ (high $\delta_3$) to high $\delta_3$ (low $\delta_2$). Relatively low values of $\varphi_c$, in the range of 32-35 degrees, produce good predictions of the stress (low values of $\delta_2$) but poor predictions in the strain (large values of $\delta_3$). The opposite is true at the largest values of $\varphi_c$, in 36-39 degrees range.

Wu \emph{et al.} \cite{Wu_A_basic_hypoplastic_constitutive_model_for_sand} have shown that the SH model oversimplifies the constitutive law in \cite{Wei_Kolymbas_Numerical_testing_of_the_stability_criterion_for_hypoplastic_constitutive_equations} and yields an unrealistic relation between the initial Poisson ratio $\nu_i$ and the friction angle at critical state $\varphi_c$. We here further explore (and corroborate) the reasons for Wu \emph{et al.} \cite{Wu_A_basic_hypoplastic_constitutive_model_for_sand}'s concerns. Figure \ref{fig:PoissonPareto} plots the initial Poisson ratio $\nu_i$, computed a posteriori from the calibrated curves, and the GA calibrated critical state $\varphi_c$ for the 400 choices of weights that lead to the Pareto front identification. The different initial conditions (see Table \ref{tab:TxD_in}) are made visible using different markers for the initial mean pressure $p_0$ and a colour scale for the void ratios $e_0$. The analysis confirms Wu \emph{et al.} \cite{Wu_A_basic_hypoplastic_constitutive_model_for_sand}'s observation on the link between $\nu_i$ and $\varphi_c$ and complements it with the critical role of the initial conditions: the mean pressure $p_0$ becomes particularly important for low values of void ratios.

Moreover, the analysis confirms that the SH model yields nonphysical negative values for $\nu_i$ at large values of $\varphi_c$, while the parameter $\alpha$ appears entirely uncorrelated from $\nu_i$. The range 36-39 degrees for $\varphi_c$ is the one that yields the best model predictions in the $(\varepsilon_a,\varepsilon_v)$ plane. 
Therefore, the nonphysical values of $\nu_i$ appear to be necessary for the SH model to overcome its underlying simplification, trading the accuracy in the prediction of the initial (contractile) phases with the accurate prediction of the expansion at the largest values of $\varepsilon_v$.

\section{Conclusions}

We studied the calibration of the SH constitutive law \cite{Wolffersdorff_A_hypoplastic_for_granular_material_with_a_predefined_limit_state_surface} in the 12 OE, and 25 CD tests on the Karlsruhe sands Wichtman and Triantafyllidis \cite{Wichtmann2016-rs}. The parameter space was extensively explored using GA \cite{Mendez_GA-cal_2022}, by setting the calibration problem in the optimisation framework proposed in \cite{Mendez2021}. In this framework, the calibration involves identifying the model parameter set that optimises a global performance measure. This is a weight-ed average of model performances in the $(\sigma_v,\,e)$, $(\varepsilon_a,\,q)$ and $(\varepsilon_a,\,\varepsilon_v)$ planes so that acting on the set of weights it is possible to prioritise accuracy on one plane over the others and study how the model parameters control the model performances. 

The available data were divided into six groups. In each of these, the model was calibrated using four combinations of weights to obtain twenty-four sets of optimal parameters. 
The analysis of the average relative deviation for each combination across the full dataset confirms that the SH model can be made sufficiently versatile to interpret all the tests, but only if a different set of parameters is used for each. 
 Moreover, a calibration that gives equal weight to all planes provides parameters that perform significantly worse than those obtained when the focus is narrowed to two or even one of the planes. This confirms the need for a compromise in the calibration, particularly in the prediction accuracy for stresses and deformations in the CD tests.

We analysed the conflicting objectives in this compromise using a Pareto front analysis and identified the parameters primarily responsible for it. Contrary to the common belief that this compromise is ruled by the parameter $\alpha$ \cite{Wichtmann2016-xw,Kadlicek2022_II,Machacek2022}, it was found that the trade-off in prediction accuracy for stresses and deformations in the CD tests is primarily due to the parameter $\varphi_c$. 

As previously highlighted by Wu \emph{et al.} \cite{Wu_A_basic_hypoplastic_constitutive_model_for_sand}, this parameter is tightly linked to the initial Poisson ratio $\nu_i$, which becomes unrealistically close to zero and eventually negative at the large values of $\varphi_c$ required to obtain good prediction in the $(\varepsilon_a,\,\varepsilon_v)$ plane. Furthermore, it was shown that the initial conditions strongly influence the relation between $\varphi_c$ and $\nu_i$, and negative values of $\nu_i$ are obtained for smaller $\varphi_c$ in the samples with the highest void ratio $e_0$.

In conclusion, we have shown that the SH model can be used to predict the soil response in the $(\sigma_v,\,e)$, $(\varepsilon_a,\,q)$ and $(\varepsilon_a,\,\varepsilon_v)$ planes if the focus is placed on each of these separately. Therefore, the SH model provides enough flexibility to study geotechnical problems such as slope stability or settlement prediction of a shallow foundation, which do not demand high accuracy in \emph{all} planes simultaneously. However, the model performances are inherently limited for problems challenging it in all planes, such as soil structure interaction. The inevitable compromises needed in the calibration have been tightly linked to the parameter $\varphi_c$ and thus to $\nu_i$.


\section*{Conflict of interest}
The authors declare no conflict of interest.

\appendix
\section{The SH model}\label{App}
In the formulation of SH by von Wolffersdorff \cite{Wolffersdorff_A_hypoplastic_for_granular_material_with_a_predefined_limit_state_surface}, the objective Zaremba-Jaumann \cite{Zaremba} stress tensor  $\accentset{\circ}{\mathbf{T}}$ is:

\begin{equation}
\accentset{\circ}{\mathbf{T}} = \frac{f_e\,f_b}{\Tr(\hat{\mathbf{T}}^2)}  \Bigl( F^2 \mathbf{D} + a^2\Tr(\hat{\mathbf{T}}\cdot \mathbf{D})\hat{\mathbf{T}}  +  f_d\,a\,F \bigr(\hat{\mathbf{T}}+\hat{\mathbf{T}}^*\bigl)\Vert \mathbf{D}  \Vert \Bigr)
\label{eq:Wolffersdorff}
\end{equation} where   $\Vert \, \Vert $ is the tensor norm $\Vert A \Vert=\sqrt{\Tr{(AA^T)}}$, $\hat{\mathbf{T}}^*= \hat{\mathbf{T}}-1/3 \mathbf{I}$, with  $\mathbf{I}$ the  identity tensor and $\hat{\mathbf{T}}={\mathbf{T}}/{\Tr({\mathbf{T}}})$. 


\begin{align} \label{eq:a}
&a				=\frac{\sqrt{3}(3-\sin \varphi_c)}{2\sqrt{2}\sin \varphi_c }\\
&F				=\sqrt{\frac{1}{8}\tan^2\psi+\frac{2-\tan^2\psi}{2+\sqrt{2}\tan\psi \cos 3\theta}}-\frac{\tan \psi}{2\sqrt{2}} \,,
\end{align}\,where: 
\begin{align}
&\tan \psi		=\sqrt{3}\Vert \hat{\mathbf{T}}^*\Vert \\
&\cos 3\theta	=-\sqrt{6}\,\frac{\Tr(\hat{\mathbf{T}}^{*3} )}{\bigr[ \Tr(\hat{\mathbf{T}}^{*2})]^{3/2}} \label{5a}\,\,.
\end{align} The barotropy coefficients $f_b$ and $f_e$ are:

\begin{align}
&f_b=\cfrac{\Biggr(\frac{e_{i0}}{e_{c0}}\Biggl)^\beta \,\cfrac{h_s}{n}\frac{1+e_i}{e_i}\Bigr(-\cfrac{\Tr(\mathbf{T})}{h_s}\Bigl)^{1-n}}{3+a^2-a\,\sqrt{3}\, \Biggr( \cfrac{e_{i0}-e_{d0}}{e_{c0}-e_{d0}}\Biggl)^\alpha}\,, \\
&f_e				=\Biggr(\frac{e_c}{e}\Biggl)^\beta\,. 
\end{align} The pyknotropy coefficient $f_d$ is defined as 

\begin{equation}
f_d=\Biggr(\frac{e-e_d}{e_c-e_d}\Biggl)^\alpha\,.
\label{eq:f_d}
\end{equation} The maximal ($e_d$), minimal ($e_i$) and critical ($e_c$)  void  ratio are linked by  the system:
 
\begin{equation}\label{eq:ei_ed_c_ed}
\frac{e_i}{e_{i0}}=\frac{e_d}{e_{d0}}=\frac{e_c}{e_{c0}}=\exp\Biggr[-\Biggr(\frac{-\Tr(\mathbf{T})}{h_s}\Biggl)^n\Biggl]\,.
\end{equation} The materials parameters $\mathbf{P}=\{e_{c0} ,  e_{d0} ,  e_{i0} ,  h_s ,  \phi ,  n ,  \alpha ,  \beta \}$ are:
\begin{itemize}
\item  $e_{c0}$,  $e_{d0}$  $e_{i0}$. These are, respectively, the critical the minimal and the maximal void ratios, obtained when $p_s=\Tr(\mathbf{T})=0 $ in \eqref{eq:ei_ed_c_ed}. 

 \item $h_s$ is called granular hardness, and Its increasing increases stiffness and dilatancy.

\item $\varphi_c$ is the critical friction angle at the critical state.

\item $n$ is a parameter influencing the barotropy of the soil, influencing the stiffness, peak resistance and dilatancy. 

\item $\alpha$ is the exponent in the calculation of the pyknotropy  coefficient $f_d$ and controls the dependency of peak friction angle on relative density. 
 
\item $\beta$ is  a  coefficient influencing barotropy and  pyknotropy. Increasing $\beta$   produces  an increase of the stiffness of material and in particular the shear stiffness.

\end{itemize}

The reader is referred to \cite{Herle_Gudehus_Determination_of_parameters_of_a_hypoplastic_constitutive_model_from_properties_of_grain_assemblies,Masin_The_influence_of_experimental_and_sampling_uncertainties_on_the_probability_of_unsatisfactory_performance_in_geotechnical_applications,Mendez2021} for more details on these parameters, and to  \cite{Mendez2021,GA_manual} for more details on the model formulation and the relevant simplifications related to the OD and CD tests.

\subsection*{Acknowledgements}
The authors gratefully acknowledge the support and the discussions with the engineer Pierantonio Cascioli, from GEINA srl, and Gabriele Sandro Toro, laboratory technician of the Department of Engineering and Geology of the Faculty Gabriele D'Annunzio of Chieti.

{\small
\bibliographystyle{ieee_fullname}
\bibliography{Bibliography}
}


\end{document}